\documentclass[twocolumn, twocolappendix]{aastex631}
\usepackage{hyperref}  
\usepackage{multirow}  
\usepackage{amsmath}

\begin{document}

\title{Magnetic Fields in the Pillars of Creation}

\author[0009-0000-7137-0066]{Adwitiya Sarkar}
\affiliation{Department of Astronomy, University of Illinois, 1002 West Green St, Urbana, IL 61801, USA}
\correspondingauthor{Adwitiya Sarkar}
\email{as155@illinois.edu}

\author[0000-0002-4540-6587]{Leslie W. Looney}
\affiliation{Department of Astronomy, University of Illinois, 1002 West Green St, Urbana, IL 61801, USA}

\author[0000-0002-7269-342X]{Marc W. Pound}
\affiliation{University of Maryland, Department of Astronomy, College Park, MD 20742-2421, USA}

\author[0000-0002-7402-6487]{Zhi-Yun Li}
\affiliation{University of Virginia, 530 McCormick Rd., Charlottesville, Virginia 22904, USA}
\affiliation{Virginia Institute for Theoretical Astronomy, University of Virginia, Charlottesville, VA 22904, USA}

\author[0000-0003-3017-4418]{Ian W. Stephens}
\affiliation{Department of Earth, Environment, and Physics, Worcester State University, Worcester, MA 01602, USA}
\affiliation{Center for Astrophysics $\vert$ Harvard \& Smithsonian, 60 Garden Street, Cambridge, MA 02138, USA}

\author[0000-0001-5811-0454]{Manuel Fern\'{a}ndez-L\'{o}pez}
\affiliation{Instituto Argentino de Radioastronom{\'i}a, CCT-La Plata (CONICET), C.C.5, 1894, Villa Elisa, Argentina}
\affiliation{Facultad de Ciencias Astronómicas y Geof{\'i}sicas, Universidad Nacional de La Plata, Argentina}

\author[0000-0002-0859-0805]{Simon Coudé}
\affiliation{Department of Earth, Environment, and Physics, Worcester State University, Worcester, MA 01602, USA}
\affiliation{Center for Astrophysics $\vert$ Harvard \& Smithsonian, 60 Garden Street, Cambridge, MA 02138, USA}

\author[0000-0001-7233-4171]{Zhe-Yu Daniel Lin}
\affiliation{Earth and Planets Laboratory, Carnegie Institution for Science, 5241 Broad Branch Rd. NW, Washington, DC 20015, USA}

\author[0000-0002-8537-6669]{Haifeng Yang}
\affiliation{Institute for Astronomy, School of Physics, Zhejiang University, Hangzhou, 310027 Zhejiang, China}

\author{Reid Faistl}
\affiliation{Department of Astronomy, University of Illinois, 1002 West Green St, Urbana, IL 61801, USA}




\begin{abstract}

Due to dust grain alignment with magnetic fields, dust polarization observations of far-infrared emission from cold molecular clouds are often used to trace  magnetic fields, allowing a probe of the effects of magnetic fields on the star formation process. We present inferred magnetic field maps of the  Pillars of Creation region within the larger M16 emission nebula, derived from dust polarization data in the 89 and 154 $\mu$m continuum using SOFIA/HAWC+. We derive magnetic field strength estimates using the Davis-Chandrasekhar-Fermi method. We compare the polarization and magnetic field strengths to column densities and dust continuum intensities across the region to build a coherent picture of the relationship between star forming activity and magnetic fields in the region. The projected magnetic field strengths derived are in the range of $\sim$50-130 $\mu$G, which is typical for clouds of similar \textit{n}($\rm{H}_{2}$), i.e., molecular hydrogen volume density on the order of 10$^{4}$-10$^{5}$ cm$^{-3}$. We conclude that star formation occurs in the finger tips when the magnetic fields are too weak to prevent radial collapse due to gravity but strong enough to oppose OB stellar radiation pressure, while in the base of the fingers the magnetic fields hinder mass accretion and consequently star formation. We also support an initial weak field model ($<$50$\mu$G) with subsequent strengthening through realignment and compression, resulting in a dynamically important magnetic field.

\end{abstract}



\section{Introduction} 
\label{sec:intro}

Magnetic fields likely play an important role in the star formation process \citep[e.g.,][]{Crutcher2012}. Comparisons of the magnetic fields with the  filamentary or elongated structures in star forming regions on the large scale have shown that the inferred magnetic fields tend to be more parallel to low-density structures and more perpendicular to high-density  structures \citep[e.g.,][]{Planck2016}.  

This generally agrees with magnetohydrodynamic simulations of super-Alfv\'enic models that predict the transition from parallel alignment to perpendicular alignment occurs at column densities below \textit{N}$_{\rm{H}}$ $\sim$ 10$^{22.8}$ cm$^{-2}$ \citep{Soler2013}. Observations using the High-resolution Airborne Wideband Camera (HAWC+) instrument mounted on the Stratospheric Observatory For Infrared Astronomy (SOFIA) toward L1688 showed a transition at a molecular hydrogen volume density of $\sim$10$^{21.7}$ cm$^{-2}$ \citep{Lee2021}. Observations of the magnetic field in elongated structures in other environments may be able to shed light into the star formation process.

One of the more well known elongated structures in star formation is the Pillars of Creation, which constitute a small region within the Eagle Nebula, also called M16, spanning about 3.6 by 4 arcminutes  \citep[approximately 6 by 6 light years or 1.8 by 1.8 pc, with our adopted distance of 1.74 kpc from][]{Kuhn2019}.

The Pillars themselves are remnants of photoevaporation due to the massive stars in the cluster NGC 6611 located to the northwest of the Pillars \citep[e.g.,][]{Thompson2002}. The heads of the Pillars are more dense than the rest of the region \citep[e.g.,][]{Karim2023}, likely due to the dense tops of the columns `protecting' the bases, thus sculpting them into the familiar elephant trunk-like structures \citep[e.g.,][]{Oliveira2008,Sugitani2002,White1999}.

\begin{figure*}
  \centering
\includegraphics[angle=0,width=0.95\textwidth]{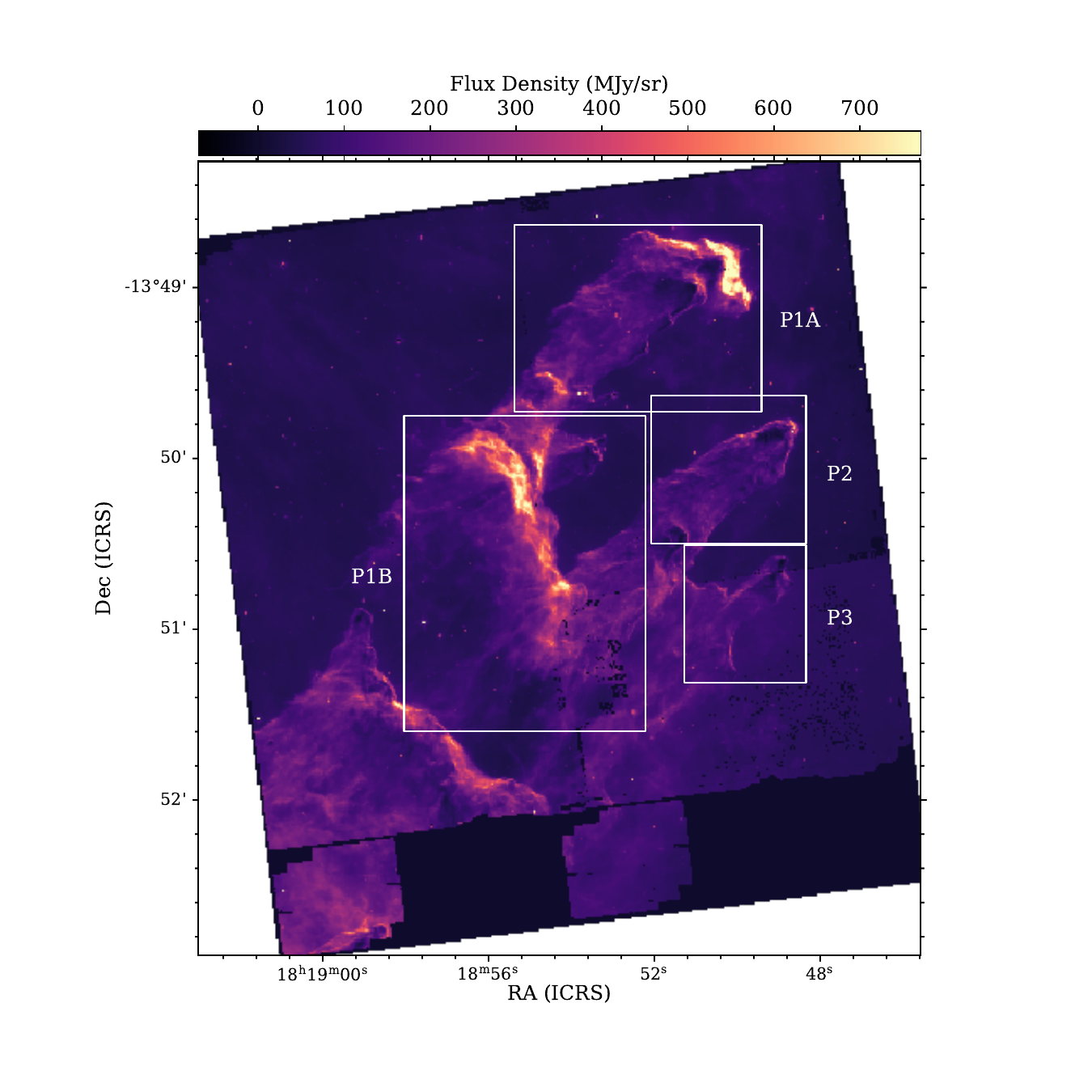}
\centering
  \caption{The regions selected for analysis: P1A, P2, P3 and P1B overlaid on JWST mid-infrared MIRI filter F770W observations.\footref{foot1}}
  \label{Hawc0}
\end{figure*}

Following \cite{McLeod2015} and \cite{Karim2023}, we divide the Pillars of Creation into four broad regions: P1A, P2, P3 and P1B. These selections are shown and labeled in Figure \ref{Hawc0}, with the regions overlaid on the 2023 JWST mid-infrared image of the Pillars taken using MIRI filter F770W\footnote{Proposal title: JWST Cycle 1 Outreach Campaign. Observation ID: V02739002001P0000000002107. PI: Klaus M. Pontoppidan. \label{foot1}}. The three large filamentary structures are called P1A, P2 and P3, from east to west. The region farther south, near the base of the fingers, is called P1B. In this paper, we shall use the term `fingers' when referring only to the three main filaments, and `Pillars' when referring to the entire region.

The Pillars are  arranged in three loosely defined `layers' in order of distance from the observer, demarcated by whether the structure lies in front of or behind NGC 6611 \citep{Pound1998, McLeod2015}, with NGC 6611 being the central layer. P1A is in the farthest layer, with its head pointing away from us. P2 lies with its tail closest to us in front of the cluster, and the head behind it. P3 lies in front of the cluster with its head pointing towards the observer. Finally, P1B also lies in front of the cluster, but with its tail pointing towards us. 
\citep[See also Figure 10 of][]{Karim2023}

Photoevaporation of the natal molecular cloud has created these structures with the dense cores at their heads shielding the bases from being blown into the surrounding ionized HII region \citep[e.g.,][]{Oliveira2008,Sugitani2002,White1999}. On a smaller scale within the fingers, there are globule substructures created in much the same way, called Evaporating Gaseous Globules or EGGs \citep[e.g.,][]{Hester1996}. About 20\% of these EGGs contain Young Stellar Objects (YSOs) \citep{McCaughrean2002}, many of which are located near the heads of P1A and P2 \citep[e.g.,][]{McCaughrean2002, Thompson2002, Sugitani2002}.

In the cold environments of such molecular clouds, the short axis of elongated dust grains are often aligned to the magnetic field direction via radiative torques \citep[e.g.,][]{Hoang2021}. Thus, dust polarization offers a probe of magnetic fields \citep[e.g.,][]{Crutcher2012}. Rotating the polarization vectors by 90$^\circ$ gives the magnetic field vectors projected on the plane-of-sky. The low dust temperatures in clouds means that they emit in the far-infrared, allowing for the use of FIR polarimeters such as SOFIA/HAWC+ \citep{Dowell2010, Harper2018} in imaging the dust polarization in these regions, which can then be used to derive magnetic field maps.

In this paper, we present SOFIA/HAWC+ polarization observations of the dust continuum in the Pillars of Creation. 
In \S \ref{sec:obs}, we present the observations and data reduction, compare the B-field morphologies in different regions, and show the results in \S \ref{sec:results}; in \S \ref{sec:da}, we analyze our reduction and compare the polarization fractions with the dust continuum intensity. We further use these regions separately to analyze their polarization angles and column densities derived from Herschel data, as well as to utilize the Davis-Chandrasekhar-Fermi method \citep{C&F1953} to find their respective magnetic field strengths. In \S \ref{sec:dis}, we examine these magnetic fields in tandem with existing theories of star formation in the region and discuss the implications. Finally, in \S \ref{sec:conc}, we present our conclusions. 

\section{Observations} \label{sec:obs}

We use Level 0 data in Bands C, D and E observed by SOFIA/HAWC+ using the chop-nod method. We give the wavelengths and angular resolutions associated with each Band in Table \ref{tab:tab0}. The observations were taken from two separate proposal IDs (PIDs), 05\_0112 and 06\_0059 (PI: Marc Pound.) For PID 05\_0112, the total exposure time was 22.2 minutes, and for PID 06\_0059 it was 56.1 minutes. The quality assessments described in the metadata for the SOFIA proposals were used to select usable data; for Band D, some files were not used in the reductions due to incomplete half width plate (HWP) angle sets. 

We use the SOFIA HAWC+ Data Reduction Pipeline v3.0.1 \citep{Clarke2023} to reduce the Level 0 data taken from the IRSA archive. Compared to the pipeline version used to reduce the M16 data in the IRSA archive, the new pipeline does a better job of noise handling, resulting in more robust processing and reduction of data. It is also more accurate in choosing usable data for processing, which can lead to different numbers of vectors while plotting inferred B-field maps post-reduction. Using this processed data, we use our own Python scripts to plot the inferred B-fields. The parameters are kept the same as those used by the pipeline to plot images, as follows: $p$/$\sigma_p$ $>$ 3.0, 0.0 $<$ $p$($\%$) $<$ 50.0 and I/$\sigma_{\rm{I}}$ $>$ 200.0. Here, $p$ is debiased polarization fraction, $p$($\%$) is debiased polarization percent, and I is Stokes I. 

\begin{table*}
    \centering
    \begin{tabular}{c c c c c}
    \hline
         & \textbf{Wavelength} & \textbf{Beam Size} & \textbf{Pixel Size} & \textbf{Mean RMS Stokes I}\\
         \hline
    \textbf{Band} & $\mu$m & arcsec (FWHM) & arcsec & mJy/$\rm{arcsec}^{2}$\\
         \hline \hline
    Band C  & 89 & 7.8 & 4.02 & 0.130 \\
    Band D   & 154 & 13.6 & 6.90 & 0.066 \\
    Band E   & 214 & 18.2 & 9.37 & 0.250 \\
   \hline
    \end{tabular}
    \caption{Wavelengths, beam sizes, pixel sizes and mean RMS Stokes I values associated with Bands C, D and E. Note that Band C and D are polarization observations while Band E is not. Mean RMS values of Bands C and D are used for the Stokes I contours in Figures \ref{Hawc1} and \ref{Hawc5}.}
    \label{tab:tab0}
\end{table*}

\section{Results} \label{sec:results}

\subsection{Dust Continuum}

Figure \ref{Hawc1} presents the Band C and D continuum emission as a colormap.
Dust continuum emission generally follows the Pillars' morphology, although we are sensitive to larger-scale dust emission surrounding the Pillars too, as seen in both maps. A similiar morphology is seen in Figure \ref{Hawc5} where both the JWST observations and the HAWC+ dust continuum emissions are overlaid. The brightness trend in the far-infrared continuum is similar to that seen in Figure \ref{Hawc0}: P1A is the brightest finger indicating either larger temperature or density. In Figure \ref{Haw} of the Appendix, we show the dust continuum observations of the larger region using the non-polarization HAWC+ observations in Band E.  

\subsection{Inferred Magnetic Field}

\begin{figure*}
  \centering
    \includegraphics[angle=0,width=1\textwidth]{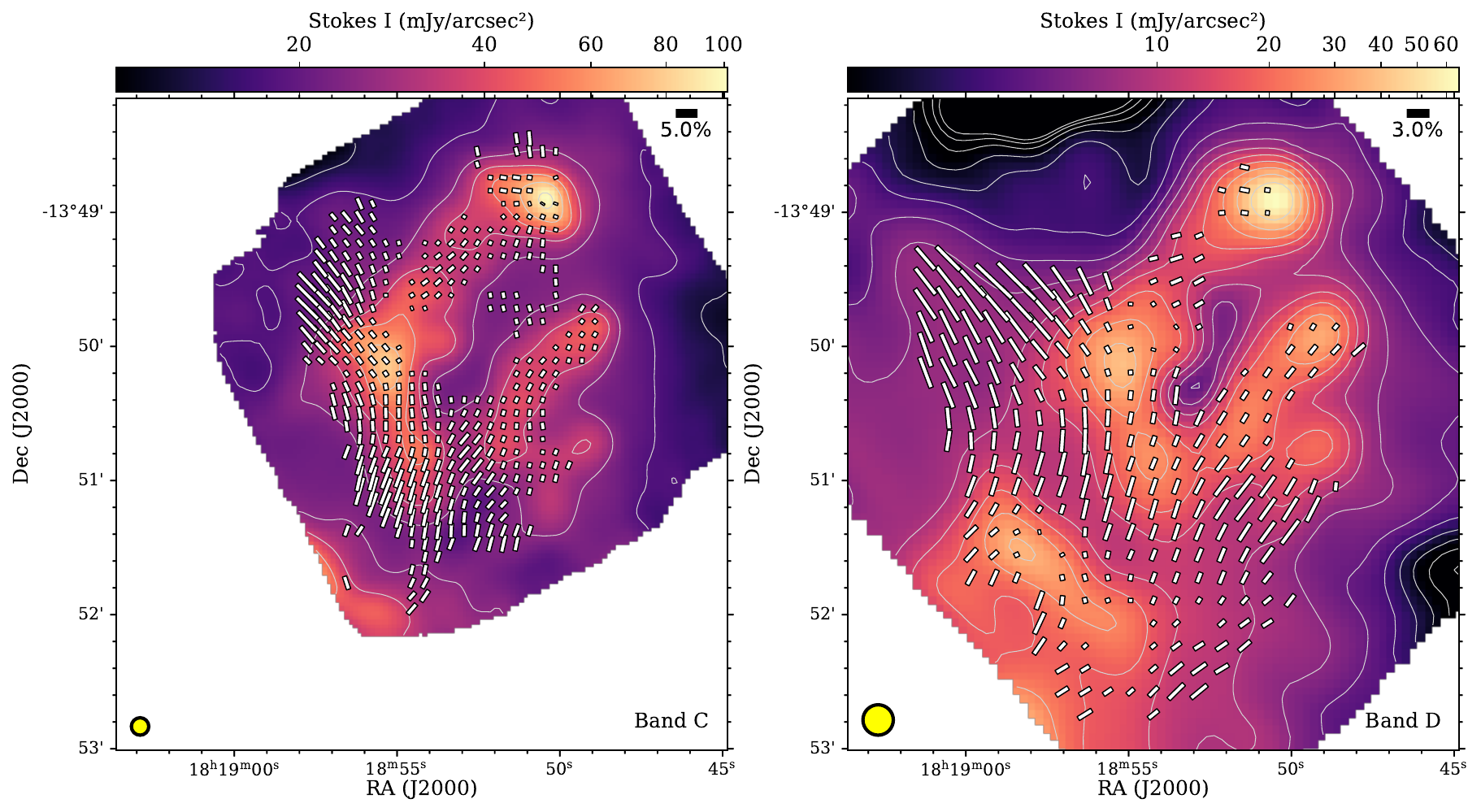}
    \centering
  \caption{Inferred magnetic field maps from the SOFIA/HAWC+ dust polarization data in Band C (left) and Band D (right). Maps are shown on the same scale. Beam sizes are shown in the bottom left corner and the polarization scale bars are shown in the top right corner. The contour levels are [10, 13.89, 19.31, 26.83, 37.28 and 51.79] for Band C and [10, 13.89, 19.31, 26.83, 37.28, 51.79, 71.97 and 100] for Band D, scaled to the mean RMS values of Stokes I in each Band, as given in Table \ref{tab:tab0}.}
  \label{Hawc1}
\end{figure*}

\begin{figure*}
  \centering
\includegraphics[angle=0,width=1\textwidth]{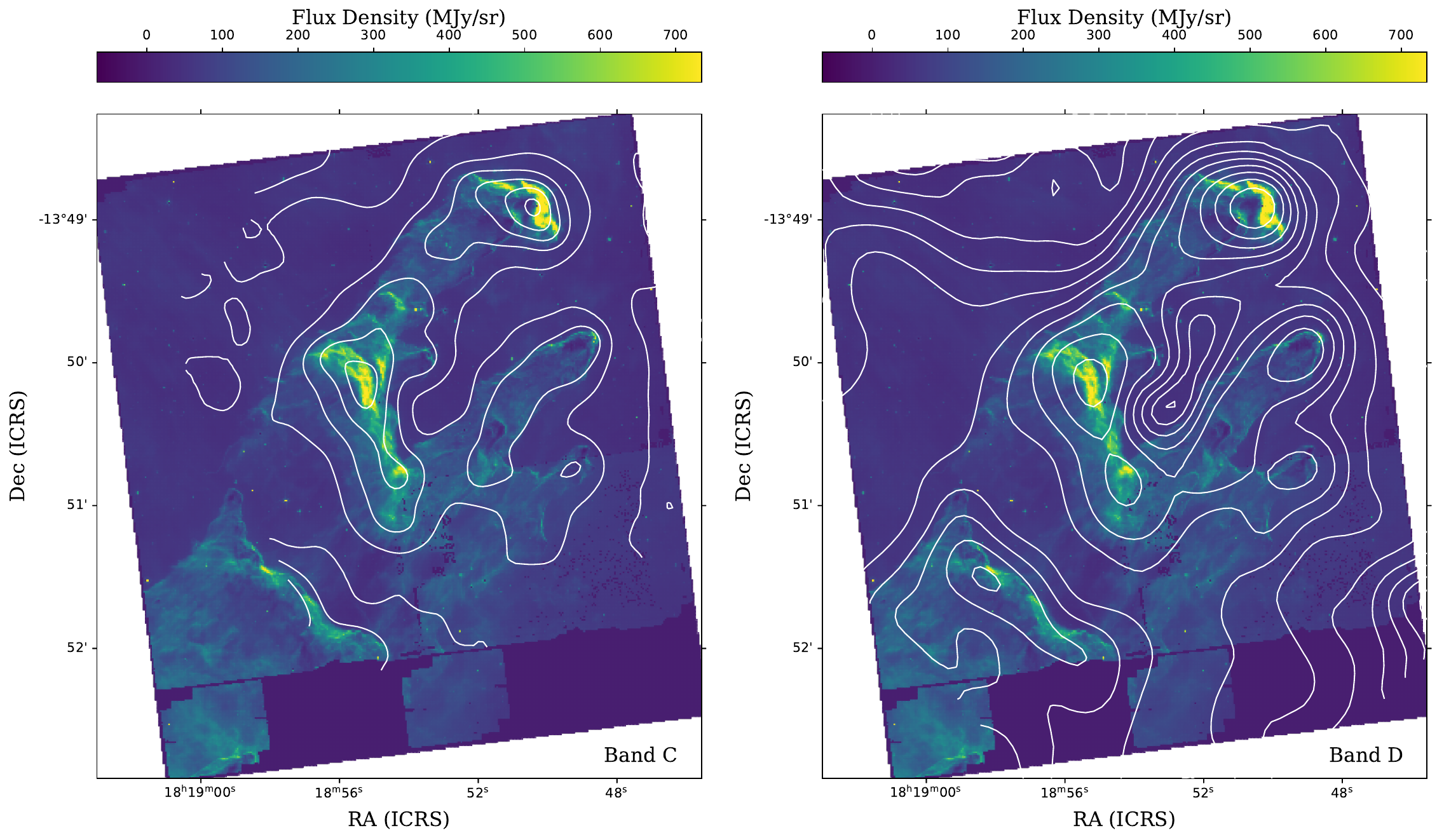}
  \caption{Stokes I contours as described in Figure \ref{Hawc1} overlaid on the JWST mid-infrared MIRI filter F770W observations from Figure \ref{Hawc0}}.
  \label{Hawc5}
\end{figure*}

Figure \ref{Hawc1} also presents the inferred B-field 
\citep[i.e., polarization rotated by 90$^\circ$; e.g.,][]{Andersson2015}, shown using half-vectors (henceforth referred to just as vectors for simplicity), meaning that the direction of the vectors are not specified. We used Nyquist beam sampling for vector visualization.

\begin{figure*}
    \includegraphics[angle=0,width=1.2\textwidth]{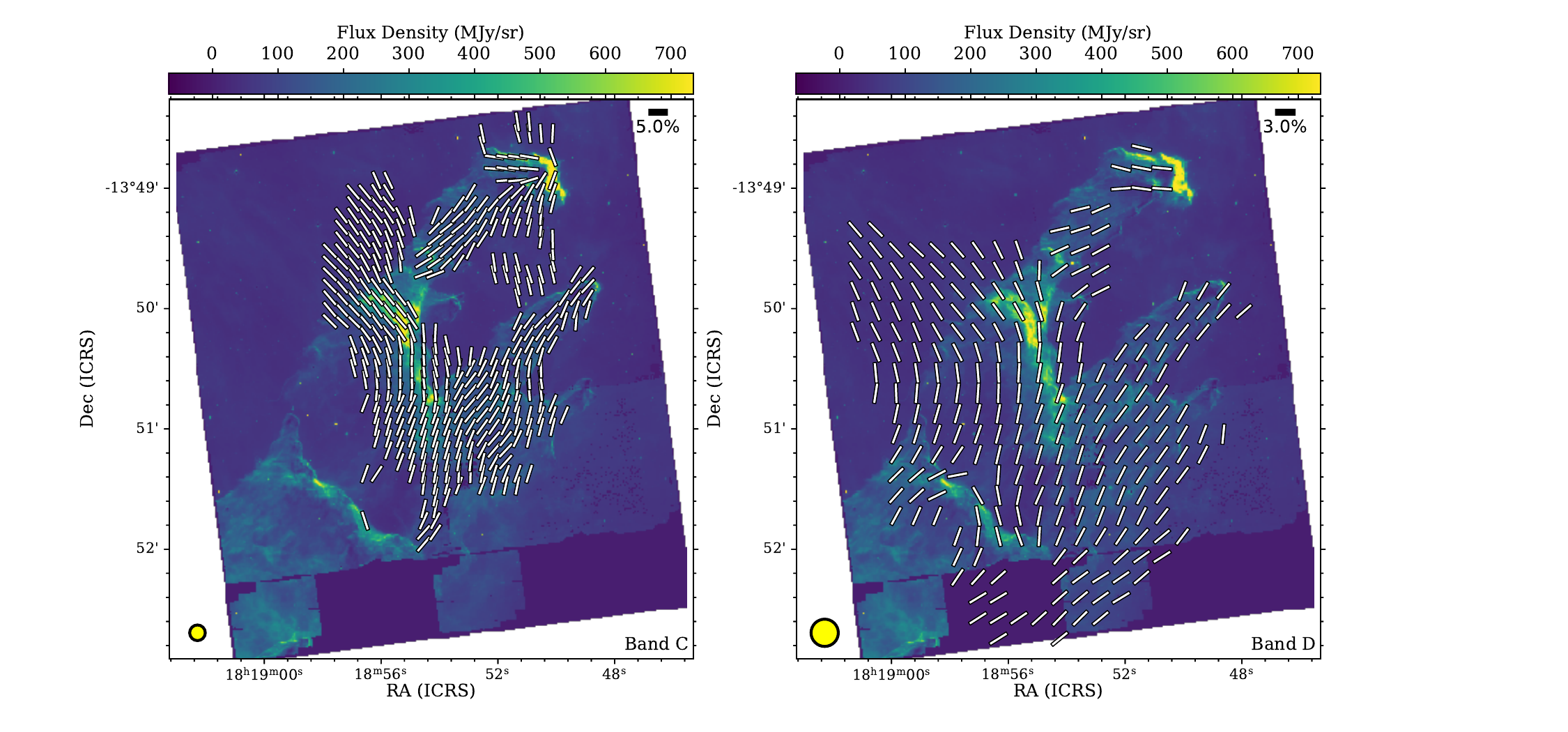}
    \caption{B-field maps as shown in Figure \ref{Hawc1} overlaid on the JWST images from Figure \ref{Hawc0}, using equal length vectors.}
  \label{Hawc2}
\end{figure*}

These maps show that the B-fields broadly align with the fingers. This trend is especially visible in Figure \ref{Hawc2}, where the unscaled magnetic field vectors from Figure \ref{Hawc1} are overlaid on the JWST image of the region. From the maps, it is clear that the inferred B-field morphologies  are generally consistent between the two Bands. 
In both Bands, the projected magnetic field extends to the northeast, outside of the filament P1A and base P1B.

In Band D, it is evident that the B-field vectors along P1A, P2 and P3 are generally aligned with the fingers, as might be expected for elongated structures of lower density \citep[e.g.][]{Planck2016}. Along P1B, the B-field curls towards the north west near the southern end of the ridge, below P2 and P3, while near the northern end of the ridge, below P1A, it bends toward the northeast, where it extends further. It is important to note that P1B is a separate structure to P1A and the field follows along the bright features seen in the JWST observations of Figure \ref{Hawc2}.

In Band C, the B-field morphologies along P2, P3 and P1B are similar to those in Band D. Along P1A, however, the vectors have more scatter, especially in the head of the structure (i.e. the brighter end). Generally, in the head of P1A, the vectors are oriented in line with the head where the dust continuum intensity is the brightest.

\cite{Pattle2018} derives a similar B-field morphology from 850 $\mu$m observations using James Clerk Maxwell Telescope's (JCMT) Submillimetre Common-User Bolometer Array 2 (SCUBA-2) POL-2 polarimeter for the B-fields In STar-forming Region Observations (BISTRO) program. Our B-field vector field is relatively smoother, but their vectors still follow the same pattern of being roughly parallel to the fingers.

Furthermore, \cite{Sugitani2007} estimates a similar orientation for the magnetic fields using near infrared polarization observations done with the South African Astronomical Observatory's (SAAO) InfraRed Survey Facility (IRSF) telescope using the Simultaneous 3-color (JHKs) InfraRed Imager for Unbiased Survey (SIRIUS) camera with the SIRIUS Polarimetry Mode (SIRPOL) instrument, arguing that the magnetic fields are aligned roughly parallel to the fingers and the direction of UV radiation from NGC 6611.

\subsection{Molecular Line Data}

\begin{figure*}
\includegraphics[angle=0,width=1.25\textwidth]{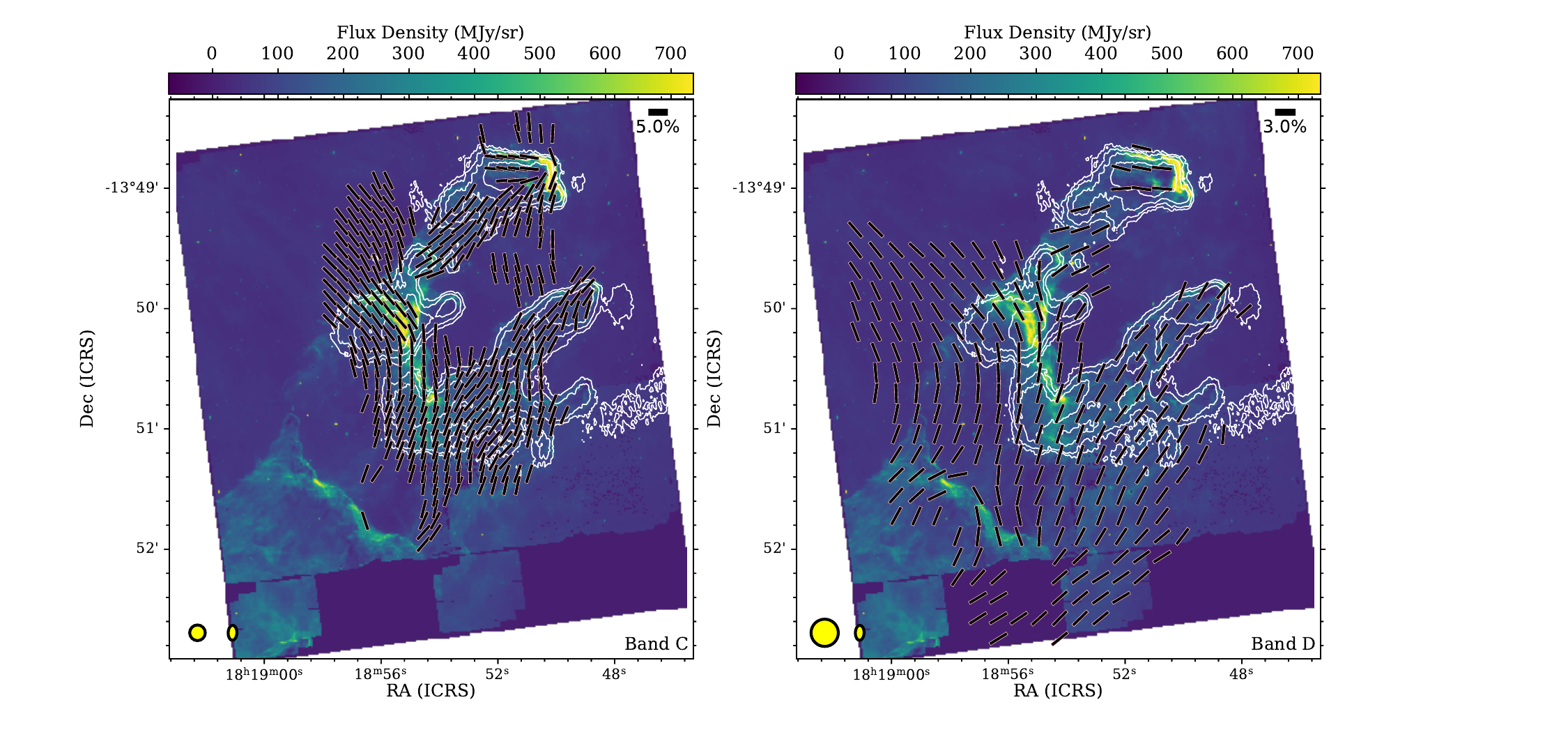}
  \caption{BIMA CO(1-0) contours overlaid on JWST image. The contour levels shown are [15.08, 25.16, 41.96, 69.99, 116.76, 194.77, 324.89, 541.95, 904.03, 1508.01] in Jy/beam $\times$ km/s. The HAWC+ and BIMA beams (left and right, respectively) are shown in the bottom left corners.}
  \label{Hawc6}
\end{figure*}

We use CO(J=1-0) data obtained from the Berkeley-Illinois-Maryland Association (BIMA) archive \citep[e.g.,][]{Pound1998} for our B-field analyses (\S \ref{sec:da22}). From Figure \ref{Hawc6}, which  shows the integrated intensity of CO(1-0), it is clear that the B-field also closely follows the contours of the gas. As expected, the dust polarization is lowest in the regions of high density (as traced by the CO).

\section{Data Analysis} \label{sec:da}

For all of our data analysis, we sample the HAWC+ polarization data with the beam FWHM (i.e., one pseudo-vector per beam given in  Table \ref{tab:tab0}) to have linearly independent pixels in our analysis.

\subsection{Polarization Efficiency and Column Densities} \label{sec:da1}

\begin{figure*}
  \centering
\includegraphics[angle=0,width=1\textwidth]{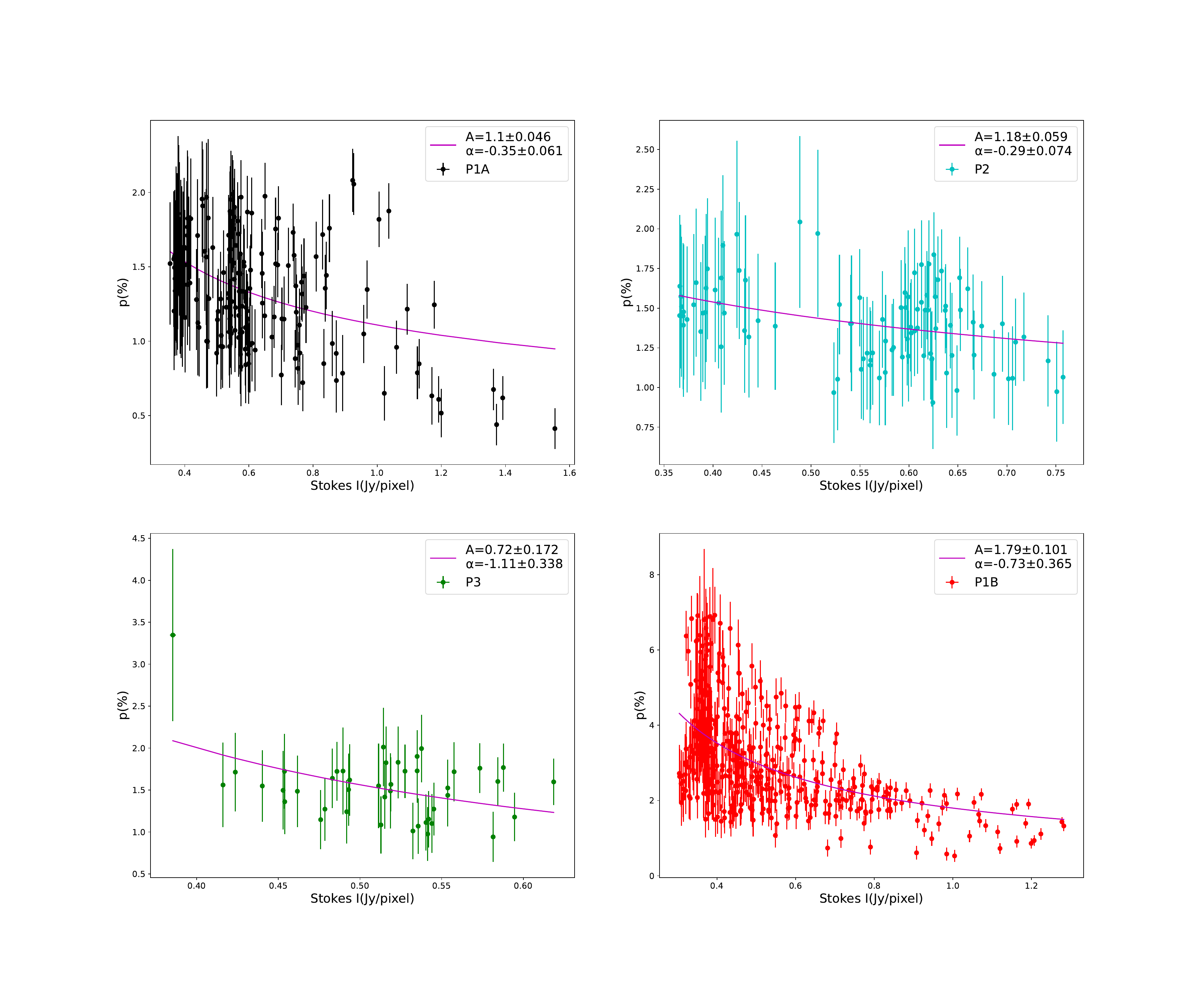}
  \caption{Debiased polarization fraction plotted against Stokes I in Band C for P1A, P2, P3 and P1B. $p$/$\sigma$$p$ $>$ 3.0, I/$\sigma$I $>$ 3.0.}
  \label{Hawc31}
\end{figure*}

\begin{figure*}
  \centering
\includegraphics[angle=0,width=1\textwidth]{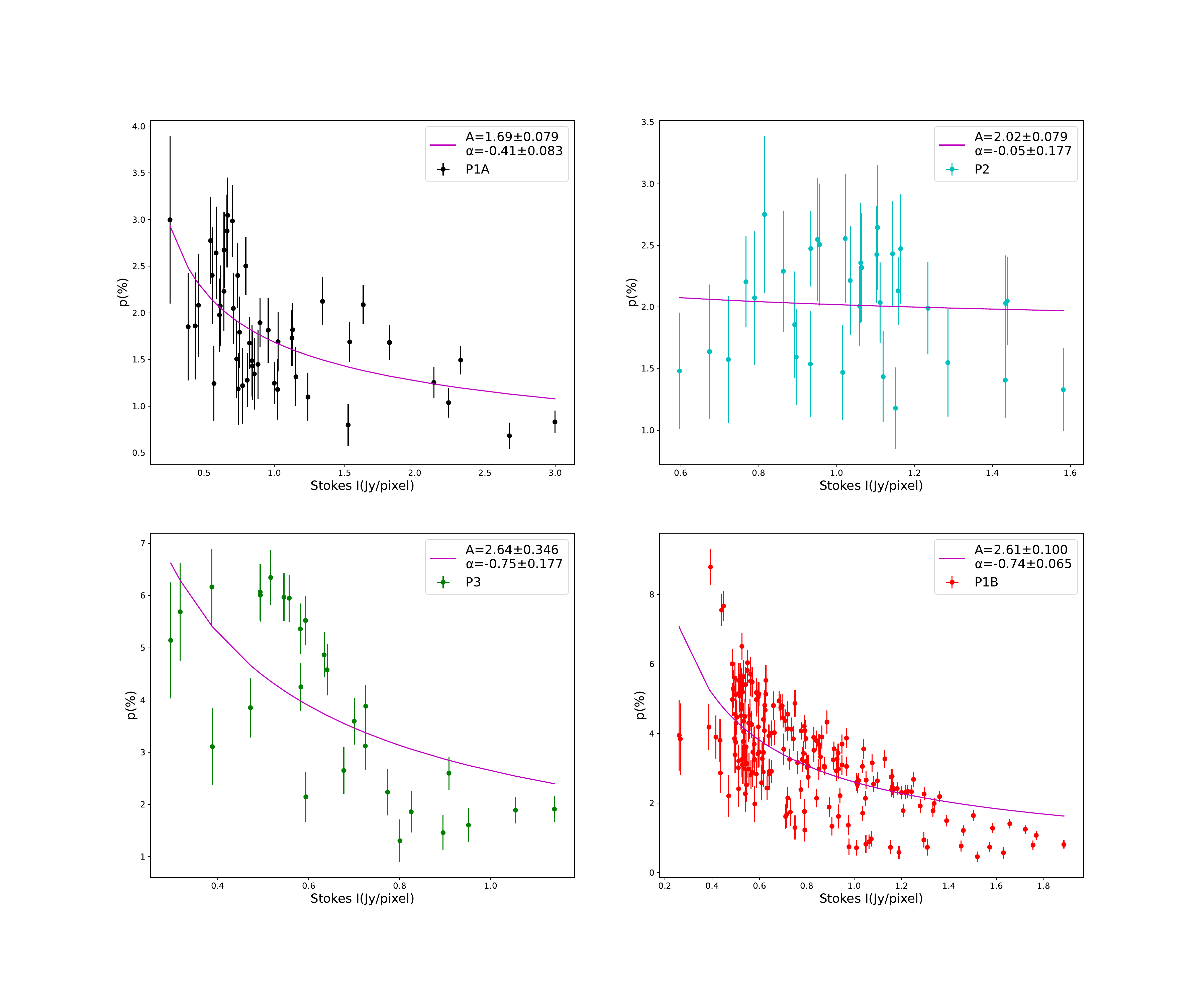}
  \caption{Debiased polarization fraction plotted against Stokes I in Band D for P1A, P2, P3 and P1B. $p$/$\sigma$$p$ $>$ 3.0, I/$\sigma$I $>$ 3.0.}
  \label{Hawc32}
\end{figure*}

Figures \ref{Hawc31} and \ref{Hawc32} show plots of the debiased polarization fraction with the Stokes I in Bands C and D of the pixels in the regions  shown in Figure \ref{Hawc0}. We use Stokes I as a general proxy for column density, although temperature, optical depth, and dust properties likely play a role too. We consider only those pixels where $p$/$\sigma_p$ $>$ 3.0 and I/$\sigma_{\rm{I}} >$ 3.0.
As is commonly observed in many regions, polarization fraction decreases inversely with Stokes I \citep[e.g.,][]{Dotson1996, Alves2014, Matthews2002}, which is thought to be related to a loss of grain alignment at higher densities due to how efficiently the dust is radiated in its environment
\citep[e.g.,][]{Alves2014,Hoang2021}. This is shown by power law fit lines in the plots, displaying lower grain alignment efficiency at higher density. We fit the following model to our data:

\begin{align} 
p={\rm{A}}\rm{I}^{\rm{\alpha}}
\end{align} \label{Eq0}

We display the values of A and $\rm\alpha$ within the plots, and we find that the reduced $\chi_{\rm{r}}^2$  $\approx$ 1. These $\rm\alpha$ values are similar to those found in other regions \citep[e.g.,][]{Soam2018, Coude2019}.

\begin{figure*}
  \centering
\includegraphics[angle=0,width=1\textwidth]{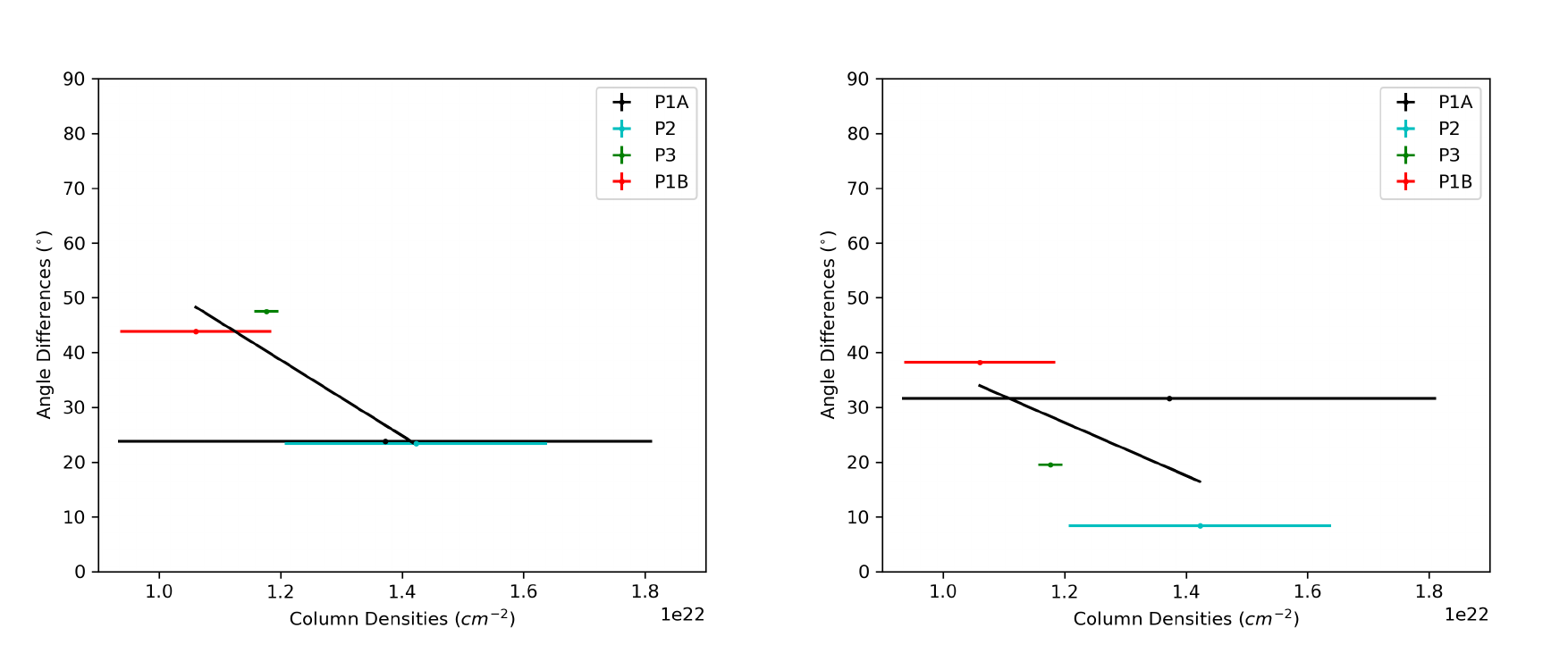}
  \caption{B-field angle differences with the structure angles in Band C (left) and Band D (right) plotted with mean $\rm{H}_{2}$ column densities for P1A, P2, P3 and P1B.
  Note that uncertainties in angle differences are plotted, but they are small since we only include the measurement uncertainties.
  }
  \label{Hawc4}
\end{figure*}

Figure \ref{Hawc4} shows B-field angle differences in both Bands plotted with mean H$_{2}$ column densities. The angle differences here are the absolute differences between the B-field vectors and the large scale structure angles or the gas flow angles relative to east of north, for which we adopt the values of 127$^{\circ}$, 140$^{\circ}$, 132$^{\circ}$ and 136$^{\circ}$ for P1A, P2, P3 and P1B, respectively \citep{Pound1998}. The dust polarization angles for these plots were found using weighted means of Stokes Q and Stokes U in each region, and subtracting 90$^\circ$ from the obtained values to get the B-field angles. The column densities were estimated by the Herschel imaging survey of OB Young Stellar Objects, or HOBYS \citep{Motte2010}. Again, we plot linear fit lines for each Band, obtaining negative slopes in both cases. Although the column densities have large uncertainties, the general trend suggests that with increasing column density, the B-fields are aligned more closely along the fingers.

\subsection{B-field Strengths} \label{sec:da2}

The Davis-Chandrasekhar-Fermi (DCF) method \citep{C&F1953}  was used to estimate the B-fields in P1A, P2 and P3. The DCF method uses the line-of-sight velocity dispersion, gas density in the cloud and the polarization angle dispersion to calculate the strength of the B-field projected on the plane-of-sky. Because polarization is observed from the projection, changes throughout a molecular cloud are not taken into account in utilizing this method, leading to some controversy surrounding it \citep[e.g.,][]{Hildebrand2009}. We use a modified version of the DCF method proposed by \cite{Crutcher2004} that utilizes a correction factor $\mathcal{Q}$
on the order of unity:

\begin{align} 
B_{\rm{POS}} = \mathcal{Q}\sqrt{4\pi\rho}\frac{\sigma_{v}}{\sigma_{\theta}} \approx 9.3\sqrt{n(\rm{H}_{2})}\frac{\Delta V}{\sigma_{\theta}} [\mu {\rm{G}}]
\end{align} \label{Eq1}

Here, $\rho$ is the gas density associated with the cloud, ${n(\rm{H}_{2})}$ is the volume density of molecular hydrogen, ${\Delta V}$=$\sqrt{8\rm{ln}(2)}~{\sigma_{v}}$ is the velocity FWHM where ${\sigma_{v}}$ is the velocity dispersion, in km/s, and ${\sigma_{\theta}}$ is the polarization angle dispersion in degrees.

There are various proposed corrections to the DCF method; \cite{Skalidis&Tassis2021} in particular argue that the classical method may overestimate the B-field even with a correction factor, and suggest a modification that addresses this issue without a correction factor, i.e., a factor of 1. However, as discussed in \cite{Liu2021}, this factor of 1 may not be accurate if the turbulent energy does not equal the fluctuating part of the magnetic energy. As such, we opt for the more widely used method described in \cite{Crutcher2004} with  a correction factor $\mathcal{Q}$=0.5 instead \citep[e.g.,][]{Ostriker2001}.

\subsubsection{Polarization Angle Dispersions} \label{sec:da21}

The DCF method uses the polarization angle dispersion, so first we remove any large scale component.
Due to the rather simple uniform polarization morphology of the Pillars and the limited number of independent beams in the observations, we approximate a large scale angle to subtract from the B-field angles in each of P1A, P2, P3 and P4, instead of opting for a more complicated structure function. These large scale angles are the gas flow angles defined in \S\ref{sec:da1}.

We use a slightly modified version of the polarization dispersion function proposed by \cite{Hildebrand2009}, as used in \cite{Stephens2022}, using only the large-scale structure contribution factor $\Delta \phi$ and the uncertainty $\sigma_{\phi}$ associated with polarization angles:

\begin{align}
    \sigma_{\theta} = \sqrt{\Delta \phi^{2} - {\sigma_{\phi}}^{2}}
\end{align} \label{Eq2}

Here, $\Delta \phi$ is taken to be the standard deviation of the difference between the B-field angle associated with each pixel and the large-scale structure angle, and $\sigma_{\phi}$ is the median of the pixel-by-pixel uncertainties, as follows:

\begin{align}
    \Delta \phi = \theta_{p} - \theta_{s} = \frac{1}{2}\rm{atan}\frac{\rm{U}}{\rm{Q}} - \theta_{s}
\end{align} \label{Eq3}

\begin{align}
    \sigma_{\phi} = \frac{1}{2}\frac{\sigma_p}{p}
\end{align} \label{Eq4}

Note that in Equation 3, ${\rm Q}$ is Stokes Q, and not the order unity correction factor $\mathcal{Q}$ discussed in \S \ref{sec:da2}.

\subsubsection{Velocity Dispersions} \label{sec:da22}

We use two methods for finding velocity dispersions, which we use separately to derive B-field strengths.
\cite{White1999} finds values of ${\Delta{V}}$ between 1.0-2.5 km/s for P1A and between 1.2-2.2 km/s for P2 using nine different tracers. For the first method (hereafter Method 1), we adopt the average ${\Delta{V}}$=1.874$\pm$0.105 km/s for the entire region. We compare the B-field strengths derived using this method with those presented by \cite{Pattle2018}, which also adopted a similar value.

For the second method (hereafter Method 2), we average the velocity dispersion over each region using the CO moment 2 maps obtained using BIMA. We then use these mean dispersion values to derive the B-field strengths. Significant velocity gradients in the fingers \citep[e.g.,][]{Karim2023,Pound1998} mean that velocity dispersion measured in large beams can be artificially increased (i.e., Method 1). We thus use the values obtained from Method 2 for further analysis in \S \ref{sec:da3}. Method 1 is used purely for a reasonable basis for comparison with the B-field strengths obtained by \cite{Pattle2018}.

\subsubsection{$n$($\rm{H}_{2}$)} \label{sec:da23}

The volume density of molecular hydrogen is taken for each region (P1A, P2, P3 and P1B) by averaging the values over each region obtained by \cite{Karim2023}. These averages are shown in Table \ref{tab:tab2}.
\\

\subsubsection{Results} \label{sec:da24}

Using Method 1 (as defined in \S \ref{sec:da22}), the derived average magnetic field strength is around 170-440 $\mu$G across both Bands in all regions. To propagate uncertainties of $\Delta V$ in Method 1, we use the standard error of the mean, since we use the mean of the values taken from \cite{White1999}. Using method 2, the values are lower, in the range of 50-130 $\mu$G. The Method 1 and 2 values span the range typical of molecular clouds of similar column and volume densities, as discussed further in \S\ref{sec:dis}. The field strengths are shown in Table \ref{tab:tab1} for each separate region.

\begin{table*}
\begin{tabular}{c c c c c c c c c}
\hline
\multirow{2}{*}{\textbf{Region}} & \multirow{2}{*}{\textbf{Mean \textbf{\textit{n}}($\rm{\textbf{H}}_{\textbf{2}}$)}} & \multirow{2}{*}{\textbf{Mean {\textbf{$\sigma_v$}}}} & \multicolumn{2}{c}{\textbf{Mass-to-flux ratios}} & \multicolumn{2}{c}{\textbf{Alfv\'en Velocities}} & \multicolumn{2}{c}{\textbf{$P_{\rm{B}}/k_{\rm{B}}$}}\\
\cline{4-9}
& & & Band C & Band D & Band C & Band D & Band C & Band D\\
& $\rm{cm}^{-3}$ & km/s & &  & km/s & km/s & K $\rm{cm}^{-3}$ & K $\rm{cm}^{-3}$\\ 
\hline \hline
P1A & 6$\times 10^{4}$ & 0.57 & 1.6$\pm$0.7 & 1.8$\pm$0.8 & 0.6$\pm$0.3 & 0.5$\pm$0.2 & 1.13$\times 10^{6}$ & 8.22$\times 10^{5}$\\ 
P2 & 1.3$\times 10^{5}$ & 0.55 & 0.8$\pm$0.3 & 0.8$\pm$0.3 & 0.8$\pm$0.3& 0.8$\pm$0.3 & 4.55$\times 10^{6}$ & 4.96$\times 10^{6}$\\ 
P3 & 9.8$\times 10^{4}$ & 0.39 & 1.0$\pm$0.4 & 1.1$\pm$0.5 & 0.6$\pm$0.3& 0.5$\pm$0.3 & 2.09$\times 10^{6}$ & 1.65$\times 10^{6}$\\ 
P1B & 1.09$\times 10^{5}$ & 0.57 & 0.7$\pm$0.3 & 0.7$\pm$0.3 & 0.8$\pm$0.4& 0.7$\pm$0.3 & 3.84$\times 10^{6}$ & 3.21$\times 10^{6}$\\ 
\hline
\end{tabular}
\caption{Mean $n$($\rm{H}_{2}$) values used for each region \citep{Karim2023}, mean $\sigma_v$ values as described in \S \ref{sec:da22} for method 2, mass-to-flux ratios derived from B-field strengths in Bands C and D, Alfv\'en velocities derived from B-field strengths and n($H_2$) in Bands C and D, $P_{\rm{B}}/k_{\rm{B}}$ derived from B-field strengths in Bands C and D.}
\label{tab:tab2}
\end{table*}

\begin{table}
\centering
\begin{tabular}{c c c c c c }
\hline
\multirow{2}{*}{\textbf{Region}} & \multicolumn{2}{ c }{\textbf{Method 1}} & \multicolumn{2}{ c }{\textbf{Method 2}} \\ 
\cline{2-5}
& Band C & Band D & Band C & Band D \\
& $\mu$G & $\mu$G & $\mu$G & $\mu$G \\
\hline \hline
P1A & 201$\pm$26 & 171$\pm$22 & 63$\pm$28 & 53$\pm$24 \\ 
P2 & 424$\pm$39 & 443$\pm$41 & 126$\pm$41 & 131$\pm$43 \\ 
P3 & 401$\pm$40 & 367$\pm$36 & 85$\pm$38 & 76$\pm$34 \\
P1B & 374$\pm$39 & 342$\pm$35 & 115$\pm$50 & 106$\pm$46 \\
\hline
\end{tabular}
\caption{Magnetic field strengths (in $\mu$G) in selected regions, derived from polarization data in Bands C and D.}
    \label{tab:tab1}
\label{tab:hierarchical}
\end{table}

\subsection{Mass-to-flux Ratios and Alfv\'enic Velocities} \label{sec:da3}

We obtain the mass-to-flux ratios in both bands using the following method from \cite{Crutcher2004}:

\begin{align}
    \lambda = \frac{M/{\Phi}_{\rm{actual}}}{M/{\Phi}_{\rm{crit}}} = 7.6 \times 10^{-21} \frac{N(\rm{H}_2)}{B_{\rm{POS}}}
\end{align} \label{Eq5}

These ratios derived from Method 2 B-field values are shown in Table \ref{tab:tab2}. The $\lambda$ values are obtained by comparing the actual mass-to-flux ratios to the critical value, and show the relative importance of the magnetic support against the gravitational force. For P2 and P1B, the $\lambda$ values are magnetically subcritical 
\citep[i.e., less than 1, meaning the actual $M$/${\Phi}$ values are less than the critical ratio;][]{Crutcher2004}, implying that the B-fields are strong enough to prevent radial collapse, as suggested also by \cite{Pattle2018}. For P1A and P3, these values are greater than 1, meaning there may a higher tendency of collapse, leading to star formation.

Measuring the Alfvenic velocities also allows us to assess the impact of the B-field on the dynamics of the region \citep{Alfven1942}:
\begin{align}
    v_A = B_{\rm{POS}}/\sqrt{4\pi \rho}
\end{align} \label{Eq6}

Here, ${\rho}$ = 2.8$m_{\rm{H}}$$n$(${\rm{H}}_{2}$) is the gas density.

Our derived Alfv\'en velocities (using Method 2 B-field values) are shown in Table \ref{tab:tab2} and are on a similar order as those obtained previously \citep{Pattle2018, Sofue2020}. The Pillars are super Alfv\'enic; the expansion of the photoionized region is faster than the Alfv\'en velocities \citep{Williams2001, McLeod2015}. \cite{McLeod2015} in particular shows that the velocity of expansion is $\sim$ 8 km/s. This means that the B-field is not strong enough to prevent gas outflow due to photoionization.

\subsection{Magnetic Pressure} \label{sec:da4}

For each region, we derive $P_{\rm{B}}/k_{\rm{B}}$ from the magnetic pressure:

\begin{align}
    P_{\rm{B}}=B_{\rm{POS}}^{2}/8\pi,
\end{align}
and where $k_{\rm{B}}$ is the Boltzmann constant. The $P_{\rm{B}}/k_{\rm{B}}$, shown in Table \ref{tab:tab2} using the Method 2 inferred magnetic fields, fall within a range of about 8$\times$10$^{5}$ to 5$\times$10$^{6}$ K $\rm{cm}^{-3}$ with errors of 65-90$\%$ and are an order of magnitude lower than those derived by \cite{Pattle2018}. However, our values still agree with their conclusion that, because the external gravitational pressure $P_{\rm{g,external}}/k_{\rm{B}}$ \citep[as also derived by][]{Pattle2018} is approximately $<=$ $P_{\rm{B}}/k_{\rm{B}}$ our values, the magnetic pressure alongside the internal gravitational pressure helps maintain pressure equilibrium.

\section{Discussion} \label{sec:dis}

There are two principal regions of active star formation in this region, located at the heads of the P1A and P2 fingers \citep[e.g.,][]{Thompson2002,McCaughrean2002}. These are also the regions with the highest column densities, as seen from the Herschel column density map \citep{Motte2010}, and among the highest dust continuum intensities in both HAWC+ Bands. This aligns with the findings of \cite{Thompson2002}, \cite{White1999}, \cite{Pound1998} and \cite{Sugitani2002} that most of the molecular gas in the fingers is concentrated at the heads. This may explain why the third region of star formation, located in the southeast P1B \citep{White1999} is less active. \cite{Pound1998} in particular shows that the velocity gradients point away from the finger tips, i.e., parallel to the B-fields. This alignment of the B-fields along the fingers could be because the radiation contribution due to the ionized HII region in the surrounding regions is much stronger than the B-field, causing the B-field to `snap' into place along the remnants of the molecular cloud. The streamlining of material along the B-field in turn shields the fingers further from OB stellar radiation. This idea of the B-field being compressed is also discussed by \cite{Pattle2018}.

The B-field values (found using both methods, as discussed in \S \ref{sec:da22}) align with proposed formation models involving the initial weak field hypothesis \citep[e.g.,][]{Pattle2018,Mackey2011}. This implies that the magnetic field is dynamically unimportant during the early stages of the formation of the fingers, causing it to become realigned parallel to them. However, our values are greater than those predicted by \cite{Mackey2011} ($<$50$\mu$G) around the Pillars. Their B-field morphologies also do not show the vectors mostly parallel to the fingers. This points to an increase in B-field strength in the fingers after their formation.

Our magnetic field strengths derived using Method 1 (Table \ref{tab:tab2}) are consistent with those of \cite{Pattle2018} using JCMT BISTRO polarization data of the region: they find B-field strengths in the range of 170-320 $\mu$G. Our range is slightly larger because we take into account all four main regions in the Pillars, whereas \cite{Pattle2018} only use the vectors in P2 and part of the shared base for analysis. These values support the hypothesis  \cite[as previously proposed by][]{Pattle2018} that, as found by \cite{Williams2007}, B-field compression can strengthen an initial weak field after it has been realigned. This further explains the morphology of the B-field east of P1 in the region of lower column density, where the B-field is no longer parallel to the fingers. This also indicates that the projected B-field outlines the cross-section of the photoevaporative interface between the molecular cloud and the HII region \citep{Hester1996}, implying that the stratification of the flow away from the region may help in keeping the B-field direction parallel to the fingers. The higher column density in the fingers compared to the surrounding regions may also keep the B-field morphology relatively simple.

As we briefly mention in \S \ref{sec:da3}, the photoevaporative flow is super Alfv\'enic. This property further points to it being in line with the initial weak field model. As for the mass-to-flux ratios, P2 and P1B are magnetically subcritical (i.e., the ratio is less than 1) while P1A and P3 are not. The former case points to the magnetic pressure being strong enough to balance gravity and prevent contraction due to gravity \citep{Crutcher2012}, while the latter case indicates greater tendency to radial collapse and consequently star formation. This is in line with observations by \cite{McCaughrean2002}, \cite{Hester1996} and \cite{Hillenbrand1993}, as we discuss at the end of \S \ref{sec:dis}. Even within P2, the local mass-to-flux ratio could be higher, leading to the relatively higher concentration of disks \citep{Hester1996, Hillenbrand1993}. Fitting our B-field values (those found using Method 2) to a B-$n$(H$_2$) power law, we derive exponents of $\kappa$$\sim$0.4 for both Bands. While this value is closer to the $\kappa$=1/2 value \citep[as favoured by e.g.,][]{Tritsis2015, Li2015} than $\kappa$=2/3, it is worth noting that studies in favour of $\kappa$$\sim$2/3 \citep[e.g,][]{Crutcher2010} are over many orders of magnitude, unlike our data. As such, accounting for the sample size, our result for the B-$n$(H$_2$) power law might not be a useful test to compare the validity of $\kappa$=1/2 and $\kappa$=2/3. Our value of $\kappa$ being $>$ 0 could still indicate either strengthening of the B-field due to compression \citep[e.g.,][]{Tritsis2015} or small scale fluctuations. However, as we mention in \S \ref{sec:da1}, the B-fields are better aligned along the fingers in regions of higher column density, strongly suggesting realignment and compression of an initial weak field, subsequently strengthening it. This is in line with the resultant pinched geometry as discussed by \cite{Pattle2018}, along with their argument in favour of a B$\propto$$\sqrt{\rm{n}}$ model for the Pillars. This suggests, alongside the initial weak field hypothesis \citep[e.g.,][]{Mackey2011}, that the B-field increasing roughly by a factor of 2 would correspond to an $n$(H$_2$) jump by a factor of 4 due to compression.

Comparing the regions of high column density with regions of active star formation and examining the B-field vectors thus gives us a good idea of the direction of the B-fields in the region, despite the related drawback of obtaining inferred B-field maps from polarization data. The B-fields are also likely opposing the direction of OB radiation from the NGC 6611 cluster, keeping the dense cores relatively intact. This is further backed by the fact that the head of P3 (which is the finger that is most directly in line with the cluster) has experienced more erosion due to stronger radiation pressure \citep{Thompson2002}.

The B-field vectors at the head of P1A being somewhat isolated in both Bands and having a different direction might suggest that the high column density in the region affects the B-field to a greater degree; thus forcing it to move along the star forming core located there. The head of P3 also experiences a higher OB association radiation due to NGC6611 \citep{Thompson2002} which might hinder the magnetic field in the region, resulting in lower polarization percent and consequently lower values of observed B-field strengths compared to P2.

In general, the second and third fingers have higher B-field strengths than P1A. Considering that the change in wavelength and therefore in the temperature across the Bands does not affect the strength in P1B nearly as much as it does in the fingers consolidates the fact that dust polarization and B-field strengths are heavily impacted by the presence of star forming cores. Furthermore, only about 20\% of the EGGs in the Pillars contain YSOs \citep{McCaughrean2002}, albeit this is a lower limit. Most of these YSOs are concentrated in the heads of the first two fingers, particularly in P1A. This, accompanied by the fact that the column densities are higher at the finger heads, strongly suggests that the B-field sweeps material along with it, aiding in star formation near the heads while hindering it in other regions. The B-fields, having evolved to be dynamically important in preventing radial collapse, would also hinder significant mass accretion by YSOs, causing them to be uncovered by photoevaporation. This further limits the rate of star formation \citep[e.g.,][]{Hester1996, Hillenbrand1993, Thompson2002}.

The role of turbulence in structural support within the Pillars has been discussed by \cite{Karim2023}. In general, they conclude that turbulent pressure is $\sim$ 1.5 $\times$ 10$^7$ K cm$^{-3}$ in the molecular gas and $\sim$ 5 $\times$ 10$^{6}$ K cm$^{-3}$ in the atomic gas. The photoevaporative flow is not turbulent \citep[e.g.,][]{Hester1996}. \cite{Karim2023} further conclude that the total pressure is magnetically dominated in both molecular and atomic gas. However, comparing our $P_{\rm{B}}/k_{\rm{B}}$ values from \S \ref{sec:da4} to their estimation of turbulent pressure, we find that the total internal pressure is dominated by turbulent pressure in the molecular gas. At the same time, we briefly discuss in \S \ref{sec:da4}, our magnetic pressure values indicate that the B-fields prevent radial collapse, as also suggested by \cite{Pattle2018}. Comparison between the roles played by the turbulent and magnetic pressures in star formation in the Pillars is, however, outside the scope of this paper, and we encourage further analysis using these data.

\section{Conclusions} \label{sec:conc}

We analyze the magnetic fields in the Pillars of Creation region of M16 or the Eagle Nebula, and derive the B-field strengths in each of the four main filamentary structures that constitute it using HAWC+ dust polarization data. We analyze proposed formation models of the region in tandem with the evolution of the magnetic field and discuss the dynamic importance thereof. We compare these magnetic fields with the column densities and present our findings of their combined implications on star formation in the region.
The main conclusions from these observations are:

\begin{itemize}

\item 
The B-field morphologies in each individual region is very smooth and aligned parallel to the direction of `flow' of the fingers, due to realignment of the initially weak B-field. They also follow the outlines of photoevaporative outflow around the region. This indicates shepherding of material along the B-fields.

\item
The B-field strengths are on the order of 50-130 $\mu$G, which are typical strengths for molecular clouds with comparable $\rm{H}_2$ volume densities approaching the order of $10^5 \rm{cm}^{-3}$. Amongst the fingers, P2 has the highest values of B-field strengths in both Bands. These values are in line with the power-law relation described in \cite{Crutcher2012}, for instance. We add onto \cite{Pattle2018}'s argument in support of an initial weak field case (initially dynamically unimportant), concluding that our B-field and column density analyses also suggest strengthening of the B-field in the region, making it dynamically important.

\item 
We argue that there is a two-way relation between star formation and B-field strength. It is likely that the magnetic fields aid in star formation in the finger tips (particularly P3) by opposing the OB association radiation pressure. Moreover, in P1A and P3, the magnetic fields are too weak to prevent collapse leading to star formation, while in P2 and P1B they are strong enough to prevent radial collapse due to gravity. This may be overcome by a higher local mass-to-flux ratio in P2, resulting in higher rates of star formation compared to elsewhere in the base of the Pillars. Conversely, the magnetic fields hinder mass accretion by YSOs and therefore star formation along the base of the fingers and other regions in the Pillars.

\end{itemize}


L.W.L. acknowledges support from NSF AST-1910364 and NSF AST-2307844. H.Y. is supported in part by National Natural Science Foundation of China (NSFC) [12473067] and the Cyrus Tang Foundation. MWP acknowledges support from NASA/USRA NNA17BF53C. ZYL is supported in part by NASA 80NSSC20K0533, NSF AST-2307199, and the Virginia Institute for Theoretical Astronomy (VITA).
We thank Guyan Zhang and Frankie J. Encalada for their inputs in the initial analyses, and Sarah Sadavoy for help with the column density maps.

Based on observations made with the NASA/DLR Stratospheric Observatory for Infrared Astronomy (SOFIA). SOFIA is jointly operated by the Universities Space Research Association, Inc. (USRA), under NASA contract NNA17BF53C, and the Deutsches SOFIA Institut (DSI) under DLR contract 50 OK 0901 to the University of Stuttgart.

The final maps and the scripts used to make them  are available in the Illinois Data Bank \citep{illinoisdatabankIDB-6987399}. The HAWC+ data and pipeline reduction code are available from the SOFIA  archive at the NASA/IPAC Infrared Science Archive.

%

\vspace{5mm}
\facilities{SOFIA(HAWC+)}


\software{astropy \citep{2013A&A...558A..33A,2018AJ....156..123A}
          }




\bibliography{main}{}

\begin{thebibliography}{}
\expandafter\ifx\csname natexlab\endcsname\relax\def\natexlab#1{#1}\fi
\providecommand{\url}[1]{\href{#1}{#1}}
\providecommand{\dodoi}[1]{doi:~\href{http://doi.org/#1}{\nolinkurl{#1}}}
\providecommand{\doeprint}[1]{\href{http://ascl.net/#1}{\nolinkurl{http://ascl.net/#1}}}
\providecommand{\doarXiv}[1]{\href{https://arxiv.org/abs/#1}{\nolinkurl{https://arxiv.org/abs/#1}}}

\bibitem[{{Alfv{\'e}n}(1942)}]{Alfven1942}
{Alfv{\'e}n}, H. 1942, \nat, 150, 405, \dodoi{10.1038/150405d0}

\bibitem[{{Alves} {et~al.}(2014){Alves}, {Frau}, {Girart}, {Franco}, {Santos}, \& {Wiesemeyer}}]{Alves2014}
{Alves}, F.~O., {Frau}, P., {Girart}, J.~M., {et~al.} 2014, \aap, 569, L1, \dodoi{10.1051/0004-6361/201424678}

\bibitem[{{Andersson} {et~al.}(2015){Andersson}, {Lazarian}, \& {Vaillancourt}}]{Andersson2015}
{Andersson}, B.~G., {Lazarian}, A., \& {Vaillancourt}, J.~E. 2015, \araa, 53, 501, \dodoi{10.1146/annurev-astro-082214-122414}

\bibitem[{{Astropy Collaboration} {et~al.}(2013){Astropy Collaboration}, {Robitaille}, {Tollerud}, {Greenfield}, {Droettboom}, {Bray}, {Aldcroft}, {Davis}, {Ginsburg}, {Price-Whelan}, {Kerzendorf}, {Conley}, {Crighton}, {Barbary}, {Muna}, {Ferguson}, {Grollier}, {Parikh}, {Nair}, {Unther}, {Deil}, {Woillez}, {Conseil}, {Kramer}, {Turner}, {Singer}, {Fox}, {Weaver}, {Zabalza}, {Edwards}, {Azalee Bostroem}, {Burke}, {Casey}, {Crawford}, {Dencheva}, {Ely}, {Jenness}, {Labrie}, {Lim}, {Pierfederici}, {Pontzen}, {Ptak}, {Refsdal}, {Servillat}, \& {Streicher}}]{2013A&A...558A..33A}
{Astropy Collaboration}, {Robitaille}, T.~P., {Tollerud}, E.~J., {et~al.} 2013, \aap, 558, A33, \dodoi{10.1051/0004-6361/201322068}

\bibitem[{{Astropy Collaboration} {et~al.}(2018){Astropy Collaboration}, {Price-Whelan}, {Sip{\H{o}}cz}, {G{\"u}nther}, {Lim}, {Crawford}, {Conseil}, {Shupe}, {Craig}, {Dencheva}, {Ginsburg}, {VanderPlas}, {Bradley}, {P{\'e}rez-Su{\'a}rez}, {de Val-Borro}, {Aldcroft}, {Cruz}, {Robitaille}, {Tollerud}, {Ardelean}, {Babej}, {Bach}, {Bachetti}, {Bakanov}, {Bamford}, {Barentsen}, {Barmby}, {Baumbach}, {Berry}, {Biscani}, {Boquien}, {Bostroem}, {Bouma}, {Brammer}, {Bray}, {Breytenbach}, {Buddelmeijer}, {Burke}, {Calderone}, {Cano Rodr{\'\i}guez}, {Cara}, {Cardoso}, {Cheedella}, {Copin}, {Corrales}, {Crichton}, {D'Avella}, {Deil}, {Depagne}, {Dietrich}, {Donath}, {Droettboom}, {Earl}, {Erben}, {Fabbro}, {Ferreira}, {Finethy}, {Fox}, {Garrison}, {Gibbons}, {Goldstein}, {Gommers}, {Greco}, {Greenfield}, {Groener}, {Grollier}, {Hagen}, {Hirst}, {Homeier}, {Horton}, {Hosseinzadeh}, {Hu}, {Hunkeler}, {Ivezi{\'c}}, {Jain}, {Jenness}, {Kanarek}, {Kendrew}, {Kern}, {Kerzendorf}, {Khvalko}, {King}, {Kirkby}, {Kulkarni},
  {Kumar}, {Lee}, {Lenz}, {Littlefair}, {Ma}, {Macleod}, {Mastropietro}, {McCully}, {Montagnac}, {Morris}, {Mueller}, {Mumford}, {Muna}, {Murphy}, {Nelson}, {Nguyen}, {Ninan}, {N{\"o}the}, {Ogaz}, {Oh}, {Parejko}, {Parley}, {Pascual}, {Patil}, {Patil}, {Plunkett}, {Prochaska}, {Rastogi}, {Reddy Janga}, {Sabater}, {Sakurikar}, {Seifert}, {Sherbert}, {Sherwood-Taylor}, {Shih}, {Sick}, {Silbiger}, {Singanamalla}, {Singer}, {Sladen}, {Sooley}, {Sornarajah}, {Streicher}, {Teuben}, {Thomas}, {Tremblay}, {Turner}, {Terr{\'o}n}, {van Kerkwijk}, {de la Vega}, {Watkins}, {Weaver}, {Whitmore}, {Woillez}, {Zabalza}, \& {Astropy Contributors}}]{2018AJ....156..123A}
{Astropy Collaboration}, {Price-Whelan}, A.~M., {Sip{\H{o}}cz}, B.~M., {et~al.} 2018, \aj, 156, 123, \dodoi{10.3847/1538-3881/aabc4f}

\bibitem[{{Chandrasekhar} \& {Fermi}(1953)}]{C&F1953}
{Chandrasekhar}, S., \& {Fermi}, E. 1953, \apj, 118, 113, \dodoi{10.1086/145731}

\bibitem[{{Clarke} \& {Vander Vliet}(2023)}]{Clarke2023}
{Clarke}, M., \& {Vander Vliet}, R. 2023, {SOFIA-USRA/sofia\_redux: v1.3.1}, v1.3.1,  Zenodo, \dodoi{10.5281/zenodo.7632852}

\bibitem[{{Coud{\'e}} {et~al.}(2019){Coud{\'e}}, {Bastien}, {Houde}, {Sadavoy}, {Friesen}, {Di Francesco}, {Johnstone}, {Mairs}, {Hasegawa}, {Kwon}, {Lai}, {Qiu}, {Ward-Thompson}, {Berry}, {Chen}, {Fiege}, {Franzmann}, {Hatchell}, {Lacaille}, {Matthews}, {Moriarty-Schieven}, {Pon}, {Andr{\'e}}, {Arzoumanian}, {Aso}, {Byun}, {Eswaraiah}, {Chen}, {Chen}, {Ching}, {Cho}, {Choi}, {Chrysostomou}, {Chung}, {Doi}, {Drabek-Maunder}, {Dowell}, {Eyres}, {Falle}, {Friberg}, {Fuller}, {Furuya}, {Gledhill}, {Graves}, {Greaves}, {Griffin}, {Gu}, {Hayashi}, {Hoang}, {Holland}, {Inoue}, {Inutsuka}, {Iwasaki}, {Jeong}, {Kanamori}, {Kataoka}, {Kang}, {Kang}, {Kang}, {Kawabata}, {Kemper}, {Kim}, {Kim}, {Kim}, {Kim}, {Kim}, {Kim}, {Kirk}, {Kobayashi}, {Koch}, {Kwon}, {Lee}, {Lee}, {Lee}, {Li}, {Li}, {Li}, {Liu}, {Liu}, {Liu}, {Liu}, {van Loo}, {Lyo}, {Matsumura}, {Nagata}, {Nakamura}, {Nakanishi}, {Ohashi}, {Onaka}, {Parsons}, {Pattle}, {Peretto}, {Pyo}, {Qian}, {Rao}, {Rawlings}, {Retter}, {Richer}, {Rigby}, {Robitaille},
  {Saito}, {Savini}, {Scaife}, {Seta}, {Shinnaga}, {Soam}, {Tamura}, {Tang}, {Tomisaka}, {Tsukamoto}, {Wang}, {Wang}, {Whitworth}, {Yen}, {Yoo}, {Yuan}, {Zenko}, {Zhang}, {Zhang}, {Zhou}, {Zhu}, \& {B-fields In STar-forming Regions Observations (BISTRO Collaboration}}]{Coude2019}
{Coud{\'e}}, S., {Bastien}, P., {Houde}, M., {et~al.} 2019, \apj, 877, 88, \dodoi{10.3847/1538-4357/ab1b23}

\bibitem[{{Crutcher}(2004)}]{Crutcher2004}
{Crutcher}, R.~M. 2004, in The Magnetized Interstellar Medium, ed. B.~{Uyaniker}, W.~{Reich}, \& R.~{Wielebinski}, 123--132

\bibitem[{{Crutcher}(2012)}]{Crutcher2012}
{Crutcher}, R.~M. 2012, \araa, 50, 29, \dodoi{10.1146/annurev-astro-081811-125514}

\bibitem[{{Crutcher} {et~al.}(2010){Crutcher}, {Wandelt}, {Heiles}, {Falgarone}, \& {Troland}}]{Crutcher2010}
{Crutcher}, R.~M., {Wandelt}, B., {Heiles}, C., {Falgarone}, E., \& {Troland}, T.~H. 2010, \apj, 725, 466, \dodoi{10.1088/0004-637X/725/1/466}

\bibitem[{{Dotson}(1996)}]{Dotson1996}
{Dotson}, J.~L. 1996, \apj, 470, 566, \dodoi{10.1086/177888}

\bibitem[{{Dowell} {et~al.}(2010){Dowell}, {Cook}, {Harper}, {Lin}, {Looney}, {Novak}, {Stephens}, {Berthoud}, {Chuss}, {Crutcher}, {Dotson}, {Hildebrand}, {Houde}, {Jones}, {Krejny}, {Lazarian}, {Moseley}, {Tassis}, {Vaillancourt}, \& {Werner}}]{Dowell2010}
{Dowell}, C.~D., {Cook}, B.~T., {Harper}, D.~A., {et~al.} 2010, in Society of Photo-Optical Instrumentation Engineers (SPIE) Conference Series, Vol. 7735, Ground-based and Airborne Instrumentation for Astronomy III, ed. I.~S. {McLean}, S.~K. {Ramsay}, \& H.~{Takami}, 77356H, \dodoi{10.1117/12.857842}

\bibitem[{{Harper} {et~al.}(2018){Harper}, {Runyan}, {Dowell}, {Wirth}, {Amato}, {Ames}, {Amiri}, {Banks}, {Bartels}, {Benford}, {Berthoud}, {Buchanan}, {Casey}, {Chapman}, {Chuss}, {Cook}, {Derro}, {Dotson}, {Evans}, {Fixsen}, {Gatley}, {Guerra}, {Halpern}, {Hamilton}, {Hamlin}, {Hansen}, {Heimsath}, {Hermida}, {Hilton}, {Hirsch}, {Hollister}, {Hostetter}, {Irwin}, {Jhabvala}, {Jhabvala}, {Kastner}, {Kov{\'a}cs}, {Lin}, {Loewenstein}, {Looney}, {Lopez-Rodriguez}, {Maher}, {Michail}, {Miller}, {Moseley}, {Novak}, {Pernic}, {Rennick}, {Rhody}, {Sandberg}, {Sandford}, {Santos}, {Shafer}, {Sharp}, {Shirron}, {Siah}, {Silverberg}, {Sparr}, {Spotz}, {Staguhn}, {Toorian}, {Towey}, {Tuttle}, {Vaillancourt}, {Voellmer}, {Volpert}, {Wang}, \& {Wollack}}]{Harper2018}
{Harper}, D.~A., {Runyan}, M.~C., {Dowell}, C.~D., {et~al.} 2018, Journal of Astronomical Instrumentation, 7, 1840008, \dodoi{10.1142/S2251171718400081}

\bibitem[{{Hester} {et~al.}(1996){Hester}, {Scowen}, {Sankrit}, {Lauer}, {Ajhar}, {Baum}, {Code}, {Currie}, {Danielson}, {Ewald}, {Faber}, {Grillmair}, {Groth}, {Holtzman}, {Hunter}, {Kristian}, {Light}, {Lynds}, {Monet}, {O'Neil}, {Shaya}, {Seidelmann}, \& {Westphal}}]{Hester1996}
{Hester}, J.~J., {Scowen}, P.~A., {Sankrit}, R., {et~al.} 1996, \aj, 111, 2349, \dodoi{10.1086/117968}

\bibitem[{{Hildebrand} {et~al.}(2009){Hildebrand}, {Kirby}, {Dotson}, {Houde}, \& {Vaillancourt}}]{Hildebrand2009}
{Hildebrand}, R.~H., {Kirby}, L., {Dotson}, J.~L., {Houde}, M., \& {Vaillancourt}, J.~E. 2009, \apj, 696, 567, \dodoi{10.1088/0004-637X/696/1/567}

\bibitem[{{Hillenbrand} {et~al.}(1993){Hillenbrand}, {Massey}, {Strom}, \& {Merrill}}]{Hillenbrand1993}
{Hillenbrand}, L.~A., {Massey}, P., {Strom}, S.~E., \& {Merrill}, K.~M. 1993, \aj, 106, 1906, \dodoi{10.1086/116774}

\bibitem[{{Hoang} {et~al.}(2021){Hoang}, {Tram}, {Lee}, {Diep}, \& {Ngoc}}]{Hoang2021}
{Hoang}, T., {Tram}, L.~N., {Lee}, H., {Diep}, P.~N., \& {Ngoc}, N.~B. 2021, \apj, 908, 218, \dodoi{10.3847/1538-4357/abd54f}

\bibitem[{{Karim} {et~al.}(2023){Karim}, {Pound}, {Tielens}, {Tiwari}, {Bonne}, {Wolfire}, {Schneider}, {Kavak}, {Mundy}, {Simon}, {G{\"u}sten}, {Stutzki}, {Wyrowski}, \& {Honingh}}]{Karim2023}
{Karim}, R.~L., {Pound}, M.~W., {Tielens}, A. G.~G.~M., {et~al.} 2023, \aj, 166, 240, \dodoi{10.3847/1538-3881/acff6c}

\bibitem[{{Kuhn} {et~al.}(2019){Kuhn}, {Hillenbrand}, {Sills}, {Feigelson}, \& {Getman}}]{Kuhn2019}
{Kuhn}, M.~A., {Hillenbrand}, L.~A., {Sills}, A., {Feigelson}, E.~D., \& {Getman}, K.~V. 2019, \apj, 870, 32, \dodoi{10.3847/1538-4357/aaef8c}

\bibitem[{{Lee} {et~al.}(2021){Lee}, {Berthoud}, {Chen}, {Cox}, {Davidson}, {Encalada}, {Fissel}, {Harrison}, {Kwon}, {Li}, {Li}, {Looney}, {Novak}, {Sadavoy}, {Santos}, {Segura-Cox}, \& {Stephens}}]{Lee2021}
{Lee}, D., {Berthoud}, M., {Chen}, C.-Y., {et~al.} 2021, \apj, 918, 39, \dodoi{10.3847/1538-4357/ac0cf2}

\bibitem[{{Li} {et~al.}(2015){Li}, {Yuen}, {Otto}, {Leung}, {Sridharan}, {Zhang}, {Liu}, {Tang}, \& {Qiu}}]{Li2015}
{Li}, H.-B., {Yuen}, K.~H., {Otto}, F., {et~al.} 2015, \nat, 520, 518, \dodoi{10.1038/nature14291}

\bibitem[{{Liu} {et~al.}(2021){Liu}, {Zhang}, {Commer{\c{c}}on}, {Valdivia}, {Maury}, \& {Qiu}}]{Liu2021}
{Liu}, J., {Zhang}, Q., {Commer{\c{c}}on}, B., {et~al.} 2021, \apj, 919, 79, \dodoi{10.3847/1538-4357/ac0cec}

\bibitem[{{Mackey} \& {Lim}(2011)}]{Mackey2011}
{Mackey}, J., \& {Lim}, A.~J. 2011, \mnras, 412, 2079, \dodoi{10.1111/j.1365-2966.2010.18043.x}

\bibitem[{{Matthews} \& {Wilson}(2002)}]{Matthews2002}
{Matthews}, B.~C., \& {Wilson}, C.~D. 2002, \apj, 571, 356, \dodoi{10.1086/339915}

\bibitem[{{McCaughrean} \& {Andersen}(2002)}]{McCaughrean2002}
{McCaughrean}, M.~J., \& {Andersen}, M. 2002, \aap, 389, 513, \dodoi{10.1051/0004-6361:20020589}

\bibitem[{{McLeod} {et~al.}(2015){McLeod}, {Dale}, {Ginsburg}, {Ercolano}, {Gritschneder}, {Ramsay}, \& {Testi}}]{McLeod2015}
{McLeod}, A.~F., {Dale}, J.~E., {Ginsburg}, A., {et~al.} 2015, \mnras, 450, 1057, \dodoi{10.1093/mnras/stv680}

\bibitem[{{Motte} {et~al.}(2010){Motte}, {Zavagno}, {Bontemps}, {Schneider}, {Hennemann}, {di Francesco}, {Andr{\'e}}, {Saraceno}, {Griffin}, {Marston}, {Ward-Thompson}, {White}, {Minier}, {Men'shchikov}, {Hill}, {Abergel}, {Anderson}, {Aussel}, {Balog}, {Baluteau}, {Bernard}, {Cox}, {Csengeri}, {Deharveng}, {Didelon}, {di Giorgio}, {Hargrave}, {Huang}, {Kirk}, {Leeks}, {Li}, {Martin}, {Molinari}, {Nguyen-Luong}, {Olofsson}, {Persi}, {Peretto}, {Pezzuto}, {Roussel}, {Russeil}, {Sadavoy}, {Sauvage}, {Sibthorpe}, {Spinoglio}, {Testi}, {Teyssier}, {Vavrek}, {Wilson}, \& {Woodcraft}}]{Motte2010}
{Motte}, F., {Zavagno}, A., {Bontemps}, S., {et~al.} 2010, \aap, 518, L77, \dodoi{10.1051/0004-6361/201014690}

\bibitem[{{Oliveira}(2008)}]{Oliveira2008}
{Oliveira}, J.~M. 2008, in Handbook of Star Forming Regions, Volume II, ed. B.~{Reipurth}, Vol.~5, 599, \dodoi{10.48550/arXiv.0809.3735}

\bibitem[{{Ostriker} {et~al.}(2001){Ostriker}, {Stone}, \& {Gammie}}]{Ostriker2001}
{Ostriker}, E.~C., {Stone}, J.~M., \& {Gammie}, C.~F. 2001, \apj, 546, 980, \dodoi{10.1086/318290}

\bibitem[{{Pattle} {et~al.}(2018){Pattle}, {Ward-Thompson}, {Hasegawa}, {Bastien}, {Kwon}, {Lai}, {Qiu}, {Furuya}, {Berry}, \& {JCMT BISTRO Survey Team}}]{Pattle2018}
{Pattle}, K., {Ward-Thompson}, D., {Hasegawa}, T., {et~al.} 2018, \apjl, 860, L6, \dodoi{10.3847/2041-8213/aac771}

\bibitem[{{Planck Collaboration} {et~al.}(2016){Planck Collaboration}, {Ade}, {Aghanim}, {Alves}, {Arnaud}, {Arzoumanian}, {Ashdown}, {Aumont}, {Baccigalupi}, {Banday}, {Barreiro}, {Bartolo}, {Battaner}, {Benabed}, {Beno{\^\i}t}, {Benoit-L{\'e}vy}, {Bernard}, {Bersanelli}, {Bielewicz}, {Bock}, {Bonavera}, {Bond}, {Borrill}, {Bouchet}, {Boulanger}, {Bracco}, {Burigana}, {Calabrese}, {Cardoso}, {Catalano}, {Chiang}, {Christensen}, {Colombo}, {Combet}, {Couchot}, {Crill}, {Curto}, {Cuttaia}, {Danese}, {Davies}, {Davis}, {de Bernardis}, {de Rosa}, {de Zotti}, {Delabrouille}, {Dickinson}, {Diego}, {Dole}, {Donzelli}, {Dor{\'e}}, {Douspis}, {Ducout}, {Dupac}, {Efstathiou}, {Elsner}, {En{\ss}lin}, {Eriksen}, {Falceta-Gon{\c{c}}alves}, {Falgarone}, {Ferri{\`e}re}, {Finelli}, {Forni}, {Frailis}, {Fraisse}, {Franceschi}, {Frejsel}, {Galeotta}, {Galli}, {Ganga}, {Ghosh}, {Giard}, {Gjerl{\o}w}, {Gonz{\'a}lez-Nuevo}, {G{\'o}rski}, {Gregorio}, {Gruppuso}, {Gudmundsson}, {Guillet}, {Harrison}, {Helou}, {Hennebelle},
  {Henrot-Versill{\'e}}, {Hern{\'a}ndez-Monteagudo}, {Herranz}, {Hildebrandt}, {Hivon}, {Holmes}, {Hornstrup}, {Huffenberger}, {Hurier}, {Jaffe}, {Jaffe}, {Jones}, {Juvela}, {Keih{\"a}nen}, {Keskitalo}, {Kisner}, {Knoche}, {Kunz}, {Kurki-Suonio}, {Lagache}, {Lamarre}, {Lasenby}, {Lattanzi}, {Lawrence}, {Leonardi}, {Levrier}, {Liguori}, {Lilje}, {Linden-V{\o}rnle}, {L{\'o}pez-Caniego}, {Lubin}, {Mac{\'\i}as-P{\'e}rez}, {Maino}, {Mandolesi}, {Mangilli}, {Maris}, {Martin}, {Mart{\'\i}nez-Gonz{\'a}lez}, {Masi}, {Matarrese}, {Melchiorri}, {Mendes}, {Mennella}, {Migliaccio}, {Miville-Desch{\^e}nes}, {Moneti}, {Montier}, {Morgante}, {Mortlock}, {Munshi}, {Murphy}, {Naselsky}, {Nati}, {Netterfield}, {Noviello}, {Novikov}, {Novikov}, {Oppermann}, {Oxborrow}, {Pagano}, {Pajot}, {Paladini}, {Paoletti}, {Pasian}, {Perotto}, {Pettorino}, {Piacentini}, {Piat}, {Pierpaoli}, {Pietrobon}, {Plaszczynski}, {Pointecouteau}, {Polenta}, {Ponthieu}, {Pratt}, {Prunet}, {Puget}, {Rachen}, {Reinecke}, {Remazeilles}, {Renault},
  {Renzi}, {Ristorcelli}, {Rocha}, {Rossetti}, {Roudier}, {Rubi{\~n}o-Mart{\'\i}n}, {Rusholme}, {Sandri}, {Santos}, {Savelainen}, {Savini}, {Scott}, {Soler}, {Stolyarov}, {Sudiwala}, {Sutton}, {Suur-Uski}, {Sygnet}, {Tauber}, {Terenzi}, {Toffolatti}, {Tomasi}, {Tristram}, {Tucci}, {Umana}, {Valenziano}, {Valiviita}, {Van Tent}, {Vielva}, {Villa}, {Wade}, {Wandelt}, {Wehus}, {Ysard}, {Yvon}, \& {Zonca}}]{Planck2016}
{Planck Collaboration}, {Ade}, P.~A.~R., {Aghanim}, N., {et~al.} 2016, \aap, 586, A138, \dodoi{10.1051/0004-6361/201525896}

\bibitem[{{Pound}(1998)}]{Pound1998}
{Pound}, M.~W. 1998, \apjl, 493, L113, \dodoi{10.1086/311131}

\bibitem[{Sarkar \& Looney(2025)}]{illinoisdatabankIDB-6987399}
Sarkar, A., \& Looney, L. 2025, Data for Magnetic Fields in the Pillars of Creation (Sarkar et al.),  University of Illinois Urbana-Champaign, \dodoi{https://doi.org/10.13012/B2IDB-6987399_V1}

\bibitem[{{Skalidis} \& {Tassis}(2021)}]{Skalidis&Tassis2021}
{Skalidis}, R., \& {Tassis}, K. 2021, \aap, 647, A186, \dodoi{10.1051/0004-6361/202039779}

\bibitem[{{Soam} {et~al.}(2018){Soam}, {Pattle}, {Ward-Thompson}, {Lee}, {Sadavoy}, {Koch}, {Kim}, {Kwon}, {Kwon}, {Arzoumanian}, {Berry}, {Hoang}, {Tamura}, {Lee}, {Liu}, {Kim}, {Johnstone}, {Nakamura}, {Lyo}, {Onaka}, {Kim}, {Furuya}, {Hasegawa}, {Lai}, {Bastien}, {Chung}, {Kim}, {Parsons}, {Rawlings}, {Mairs}, {Graves}, {Robitaille}, {Liu}, {Whitworth}, {Eswaraiah}, {Rao}, {Yoo}, {Houde}, {Kang}, {Doi}, {Choi}, {Kang}, {Coud{\'e}}, {Li}, {Matsumura}, {Matthews}, {Pon}, {Di Francesco}, {Hayashi}, {Kawabata}, {Inutsuka}, {Qiu}, {Franzmann}, {Friberg}, {Greaves}, {Kirk}, {Li}, {Shinnaga}, {van Loo}, {Aso}, {Byun}, {Chen}, {Chen}, {Chen}, {Ching}, {Cho}, {Chrysostomou}, {Drabek-Maunder}, {Eyres}, {Fiege}, {Friesen}, {Fuller}, {Gledhill}, {Griffin}, {Gu}, {Hatchell}, {Holland}, {Inoue}, {Iwasaki}, {Jeong}, {Kang}, {Kemper}, {Kim}, {Kim}, {Lacaille}, {Lee}, {Li}, {Liu}, {Liu}, {Moriarty-Schieven}, {Nakanishi}, {Ohashi}, {Peretto}, {Pyo}, {Qian}, {Retter}, {Richer}, {Rigby}, {Savini}, {Scaife}, {Tang},
  {Tomisaka}, {Wang}, {Wang}, {Yen}, {Yuan}, {Zhang}, {Zhang}, {Zhou}, {Zhu}, {Andr{\'e}}, {Dowell}, {Falle}, {Tsukamoto}, {Kanamori}, {Kataoka}, {Kobayashi}, {Nagata}, {Saito}, {Seta}, {Hwang}, {Han}, {Lee}, \& {Zenko}}]{Soam2018}
{Soam}, A., {Pattle}, K., {Ward-Thompson}, D., {et~al.} 2018, \apj, 861, 65, \dodoi{10.3847/1538-4357/aac4a6}

\bibitem[{{Sofue}(2020)}]{Sofue2020}
{Sofue}, Y. 2020, \mnras, 492, 5966, \dodoi{10.1093/mnras/staa226}

\bibitem[{{Soler} {et~al.}(2013){Soler}, {Hennebelle}, {Martin}, {Miville-Desch{\^e}nes}, {Netterfield}, \& {Fissel}}]{Soler2013}
{Soler}, J.~D., {Hennebelle}, P., {Martin}, P.~G., {et~al.} 2013, \apj, 774, 128, \dodoi{10.1088/0004-637X/774/2/128}

\bibitem[{{Stephens} {et~al.}(2022){Stephens}, {Myers}, {Zucker}, {Jackson}, {Andersson}, {Smith}, {Soam}, {Battersby}, {Sanhueza}, {Hogge}, {Smith}, {Novak}, {Sadavoy}, {Pillai}, {Li}, {Looney}, {Sugitani}, {Coud{\'e}}, {Guzm{\'a}n}, {Goodman}, {Kusune}, {Santos}, {Zuckerman}, \& {Encalada}}]{Stephens2022}
{Stephens}, I.~W., {Myers}, P.~C., {Zucker}, C., {et~al.} 2022, \apjl, 926, L6, \dodoi{10.3847/2041-8213/ac4d8f}

\bibitem[{{Sugitani} {et~al.}(2002){Sugitani}, {Tamura}, {Nakajima}, {Nagashima}, {Nagayama}, {Nakaya}, {Pickles}, {Nagata}, {Sato}, {Fukuda}, \& {Ogura}}]{Sugitani2002}
{Sugitani}, K., {Tamura}, M., {Nakajima}, Y., {et~al.} 2002, \apjl, 565, L25, \dodoi{10.1086/339196}

\bibitem[{{Sugitani} {et~al.}(2007){Sugitani}, {Watanabe}, {Tamura}, {Kandori}, {Hough}, {Nishiyama}, {Nakajima}, {Kusakabe}, {Hashimoto}, {Nagayama}, {Nagashima}, {Kato}, \& {Fukuda}}]{Sugitani2007}
{Sugitani}, K., {Watanabe}, M., {Tamura}, M., {et~al.} 2007, \pasj, 59, 507, \dodoi{10.1093/pasj/59.3.507}

\bibitem[{{Thompson} {et~al.}(2002){Thompson}, {Smith}, \& {Hester}}]{Thompson2002}
{Thompson}, R.~I., {Smith}, B.~A., \& {Hester}, J.~J. 2002, \apj, 570, 749, \dodoi{10.1086/339738}

\bibitem[{{Tritsis} {et~al.}(2015){Tritsis}, {Panopoulou}, {Mouschovias}, {Tassis}, \& {Pavlidou}}]{Tritsis2015}
{Tritsis}, A., {Panopoulou}, G.~V., {Mouschovias}, T.~C., {Tassis}, K., \& {Pavlidou}, V. 2015, \mnras, 451, 4384, \dodoi{10.1093/mnras/stv1133}

\bibitem[{{White} {et~al.}(1999){White}, {Nelson}, {Holland}, {Robson}, {Greaves}, {McCaughrean}, {Pilbratt}, {Balser}, {Oka}, {Sakamoto}, {Hasegawa}, {McCutcheon}, {Matthews}, {Fridlund}, {Tothill}, {Huldtgren}, \& {Deane}}]{White1999}
{White}, G.~J., {Nelson}, R.~P., {Holland}, W.~S., {et~al.} 1999, \aap, 342, 233

\bibitem[{{Williams}(2007)}]{Williams2007}
{Williams}, R.~J.~R. 2007, in Astrophysics and Space Science Proceedings, Vol.~1, Diffuse Matter from Star Forming Regions to Active Galaxies - A Volume Honouring John Dyson, ed. T.~W. {Hartquist}, J.~M. {Pittard}, \& S.~A.~E.~G. {Falle}, 129, \dodoi{10.1007/978-1-4020-5425-9_7}

\bibitem[{{Williams} {et~al.}(2001){Williams}, {Ward-Thompson}, \& {Whitworth}}]{Williams2001}
{Williams}, R.~J.~R., {Ward-Thompson}, D., \& {Whitworth}, A.~P. 2001, 327, 788, \dodoi{10.1046/j.1365-8711.2001.04757.x}

\end{thebibliography}
\bibliographystyle{aasjournal}

\appendix \label{Appendix}

\begin{figure*}
  \centering
\includegraphics[angle=0,width=0.75\textwidth]{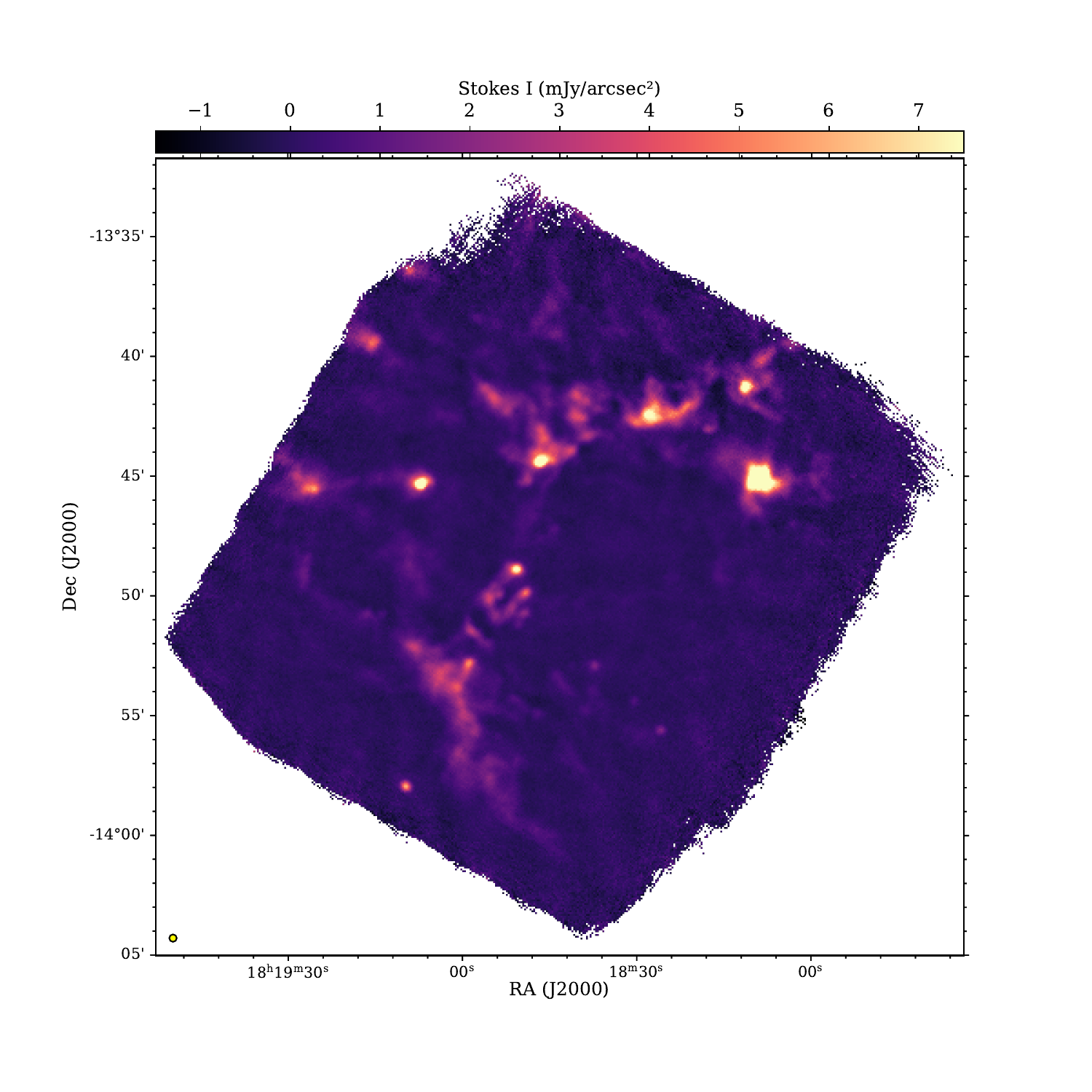}
  \caption{Band E dust continuum imaged by SOFIA/HAWC+ shows the larger region. Note that the Pillars are to the southeast of center.}
  \label{Haw}
\end{figure*}

Figure \ref{Haw} shows the dust continuum imaged in Band E in the total intensity configuration by SOFIA/HAWC+. The coverage is larger than in Bands C and D, imaged in the polarization configuration.

\end{document}